\documentclass[a4paper,11pt]{article}
\usepackage{a4wide}
\usepackage{amsmath,amssymb}
\usepackage{graphics,graphicx}
\usepackage{color}
\usepackage{cite}
\usepackage{hyperref}
\usepackage[all,cmtip]{xy}
\usepackage[T1]{fontenc} 
\usepackage[utf8]{inputenc} 
\usepackage{lmodern} 
\usepackage{multirow} 
\usepackage{tikz} 
\usetikzlibrary{arrows,calc,shadows,decorations.markings}
\usepackage{booktabs}

\newcommand{\udot}[1]{%
    \tikz[baseline=(todotted.base)]{
        \node[inner sep=1pt,outer sep=0pt] (todotted) {#1};
        \draw[densely dotted] (todotted.south west) -- (todotted.south east);
    }%
}%
\newcommand{\udash}[1]{%
    \tikz[baseline=(todotted.base)]{
        \node[inner sep=1pt,outer sep=0pt] (todotted) {#1};
        \draw[densely dashed] (todotted.south west) -- (todotted.south east);
    }%
}%

\numberwithin{equation}{section}

\begin{document}
\thispagestyle{empty}
\phantom{}
\vspace{2cm}
\begin{center}
{\LARGE\bf Hemisphere Partition Function and Monodromy}
\end{center}
\vspace{8mm}
\begin{center}
{\large David Erkinger\footnote{{\tt erkinger@hep.itp.tuwien.ac.at}},
  Johanna Knapp\footnote{{\tt knapp@hep.itp.tuwien.ac.at}}}
\end{center}
\vspace{3mm}
\begin{center}
{\em Institute for Theoretical Physics, TU Wien\\ Wiedner Hauptstrasse 8-10, 1040  Vienna, Austria}
\end{center}
\vspace{15mm}
\begin{abstract}
\noindent We discuss D-brane monodromies from the point of view of the gauged linear sigma model. We give a prescription on how to extract monodromy matrices directly from the hemisphere partition function. We illustrate this procedure by recomputing the monodromy matrices associated to one-parameter Calabi-Yau hypersurfaces in weighted projected space.
\end{abstract}
\newpage
\setcounter{tocdepth}{1}
\tableofcontents
\setcounter{footnote}{0}
\section{Introduction}
In the recent years, progress in the understanding of supersymmetric gauge theories in various dimensions has also brought new methods for computing  quantum corrections in Calabi-Yau compactifications of string theory. 

Calabi-Yaus arise as low-energy configurations, called phases, of $\mathcal{N}=(2,2)$ gauged linear sigma models (GLSMs) \cite{Witten:1993yc}. The FI-theta parameters of the gauge theory can be identified with the K\"ahler moduli of the Calabi-Yau. Using supersymmetric localization, various partition functions on curved backgrounds have turned out to compute quantities that are crucial in Calabi-Yau compactifications of type II string theory. The sphere partition function \cite{Benini:2012ui,Doroud:2012xw} has been shown to compute the exact K\"ahler potential on the quantum K\"ahler moduli space of the Calabi-Yau \cite{Jockers:2012dk,Gomis:2012wy}, and has been used to compute Gromov-Witten invariants. The elliptic genus has been computed in \cite{Benini:2013nda,Benini:2013xpa}. Supersymmetric correlation functions, including in particular the Yukawa couplings, have recently been computed from the GLSM on the deformed sphere \cite{Closset:2015rna}. The annulus partition function computes the open Witten index \cite{Honda:2013uca,Hori:2013ika}. The main focus of this work will be on the hemisphere partition function \cite{Sugishita:2013jca,Honda:2013uca,Hori:2013ika} which computes the fully quantum corrected central charges of B-type D-branes.

The moduli spaces of Calabi-Yaus have an interesting mathematical structure which in particular gives rise to non-trivial monodromies around distinguished loci in the moduli space. Using mirror symmetry, the monodromy matrices can be extracted from the period integrals. For the quintic this was already done in \cite{Candelas:1990rm}. An interpretation of the monodromy in terms of transporting D-branes along non-contractible loops in the moduli space has been given for instance in \cite{MR2700280,MR1831820,MR1852194,Mayr:2000as,Douglas:2000gi,MR2126495,Aspinwall:2001dz,Aspinwall:2002nw,Distler:2002ym,Jockers:2006sm,Herbst:2008jq,MR2923950}.

The aim of this work is to use the information encoded in the hemisphere partition function to compute monodromy matrices associated to Calabi-Yau compactifications. In this approach we compute the monodromy by transporting a set of D-branes, characterized by matrix factorizations of the GLSM superpotential along closed, non-contractible paths in the K\"ahler moduli space. In \cite{Herbst:2008jq,MR2923950} an account of D-branes and D-brane transport in the K\"ahler moduli space, including monodromies, has been given for abelian GLSMs. We will make heavy use of these results. The new observation is that the monodromy matrices can be read off directly from the integrand of the hemisphere partition function. This has been applied to the quintic in \cite{Knapp:2016rec}. Here we expand and refine this discussion and apply the method to one-parameter Calabi-Yau threefolds in weighted projective space. Thereby we will rederive the monodromy matrices that have been computed using mirror symmetry in \cite{Font:1992uk,Klemm:1992tx}. While we choose these well-studied examples for illustrative purposes, the method is in principle applicable to any Calabi-Yau that arises from a GLSM. In practice, in particular non-abelian GLSMs require more work, but this approach to compute monodromy matrices also seems to work there \cite{wip}.

The advantage of the GLSM method is that it is purely algebraic. To compute the monodromy matrices it is not required to solve differential equations. It is not even necessary to explicitly evaluate the hemisphere partition function using the residue theorem. It is also not strictly necessary to make any reference to the D-branes in the phases of the GLSM\footnote{Of course, Picard-Fuchs equations, D-brane charges obtained from evaluating the hemisphere partition function and an understanding how GLSM branes map to phases are very useful for understanding the objects we are dealing with. Therefore we will make use of all this information, but we want to stress that it would not be necessary.}. Furthermore it is possible to compute the monodromy around each distinguished locus in the moduli space independently. The group properties of the monodromy matrices are not required. Rather, one can recover them from the results. 

This article is organized as follows. In section \ref{sec-glsm} we recall the basic notions of the GLSM and give a summary of the GLSMs associated to one-parameter Calabi-Yau hypersurfaces in toric ambient spaces. Furthermore we discuss B-branes in GLSMs and give the definition of the hemisphere partition function. In section \ref{sec-branes} we summarize the main results of \cite{Herbst:2008jq} on D-brane transport in the GLSM. Of particular importance is the notion of grade restriction, which is required for transporting D-branes across phase boundaries. Grade restriction fixes concrete paths in the moduli space, and hence is crucial for the monodromy calculations. With this information at hand, we have all the ingredients to compute the monodromy in the GLSM. We then give a prescription on how to read off the monodromy matrices from the integrand of the hemisphere partition function, where we distinguish between those monodromies where a phase boundary is crossed and those where this does not happen. Section \ref{sec-onepar} is devoted to concrete constructions and explicit examples of GLSM branes as matrix factorizations of the GLSM superpotential, with focus on Calabi-Yau hypersurfaces arising as phases of $U(1)$ GLSMs. Here we can borrow a lot of technology from constructions of B-branes in geometry, Landau-Ginzburg theories and boundary CFTs. In section \ref{sec-monodromy} we finally compute the monodromy matrices for one-parameter Calabi-Yau hypersurfaces in weighted projective space by applying the steps outlined in section \ref{sec-branes}, before we end with some concluding remarks.\\\\
 This work is based on the master's thesis of D.E. \cite{erkinger}.\\\\
{\bf Acknowledgments:} We would like to thank Emanuel Scheidegger for comments on the manuscript. JK would like to thank Richard Eager, Kentaro Hori, Mauricio Romo and Emanuel Scheidegger for discussions and collaboration on related projects. JK would further like to thank the Mainz Institute for Theoretical Physics (MITP) for hospitality during the completion of this work.
\section{GLSM, B-branes and hemisphere partition function}
\label{sec-glsm}
In this section we summarize the necessary definitions of the GLSM. We further discuss the description of B-type D-branes in the GLSM in terms of matrix factorizations and give the definition of the hemisphere partition function. 
\subsection{GLSM data and phases}
To define a GLSM we choose a gauge group $G$ and a vector space $V$. We denote by $\phi_i\in V$ with $i=1,\ldots,\mathrm{dim}V$ the scalar components of the chiral matter multiplet. Further, we choose a faithful complex representation $\rho_V:G\rightarrow GL(V)$. In the Calabi-Yau case this gets restricted to $\rho_V:G\rightarrow SL(V)$. The model has a $U(1)$ R-symmetry with representation $R:U(1)\rightarrow GL(V)$. The representations $\rho_V$ and $R$ commute, and we further impose the charge integrality condition \cite{Hori:2013ika}: $R(e^{i\pi})=\rho_V(J)$ with $J\in G$. Let $T\subset G$ be a maximal torus of $G$, and $\mathfrak{t}=\mathrm{Lie}(T)$ and $\mathfrak{g}=\mathrm{Lie}(G)$ the respective Lie algebras. We denote by $Q\in \mathfrak{t}^{\ast}_{\mathbb{C}}$ the gauge charges of the chiral matter. Further, $\sigma\in \mathfrak{g}_{\mathbb{C}}$ are the scalar components of the vector multiplet. The parameters $t\in \mathfrak{g}^{\ast G}_{\mathbb{C}}$ can be decomposed as $t=\zeta-i\theta$ in terms of Fayet-Illiopoulos parameters $\zeta$ and theta angles $\theta$. In order to obtain a compact Calabi-Yau we consider models with non-vanishing gauge invariant superpotential $W(\phi)$ of R-charge $2$. Further, there is a linear twisted superpotential $\widetilde{W}(\sigma)=- t(\sigma)$.

The classical vacua are determined by the zeroes of the potential
\begin{equation}
  U=\frac{1}{8e^2}|[\sigma,\bar{\sigma}]|^2+\frac{1}{2}\left(|Q\sigma\phi|^2+|Q\bar{\sigma}\phi|^2\right)+\frac{e^2}{2}\left(\mu(\phi)-\zeta\right)^2+|dW(\phi)|^2,
  \end{equation}
where $e$ is the gauge coupling and $\mu:V\rightarrow i\mathfrak{g}^{\ast}$ is the moment map. The first term constrains $\sigma$ to take values in the Cartan subalgebra $\mathfrak{t}_{\mathbb{C}}\subset\mathfrak{g}_{\mathbb{C}}$. The latter two terms determine the D-term equations
\begin{equation}
  \mu(\phi)-\zeta=0,
  \end{equation}
and F-term equations
\begin{equation}
  dW(\phi)=0.
  \end{equation}
Depending on the value of $\zeta$, these equations can have different solutions with $\phi\neq0$ which break some or all of the gauge symmetry. The different vacua are referred to as phases of the GLSM. For the models we are about to consider, there are Higgs branches, where $\sigma=0$ and the gauge symmetry is either completely broken, or to a finite subgroup. The vacuum manifold can then be described in terms of a quotient
\begin{equation}
  \label{quotient}
 dW^{-1}(0)\cap \mu^{-1}(\zeta)/G\simeq dW^{-1}(0)\cap (V-I_{\zeta})/G_{\mathbb{C}}.
\end{equation}
$I_{\zeta}$ is called the deleted set and contains those field configurations $\phi$ for which the quotient is ill-defined. 

On the Coulomb branch, which exists for particular values of $\zeta$ at the boundary between two phases, the classical vacuum is $\phi=0$ and $\sigma\neq0$. The gauge group is broken to its maximal commuting subgroup, and $\sigma$ can take arbitrary values in $\mathfrak{t}_{\mathbb{C}}$. 
Quantum corrections generate an effective potential for $\sigma$:
\begin{equation}
  \widetilde{W}_{eff}(\sigma)=\widetilde{W}(\sigma)+\pi i\sum_{\alpha>0} \alpha(\sigma)-\sum_{Q} Q(\sigma) (\log Q(\sigma)-1),
  \end{equation}
where $\alpha$ denotes the positive roots of $\mathfrak{g}$. The critical set $t_{eff}=-\mathrm{d}\widetilde{W}_{eff}$ determines an effective FI-theta parameter and generates an effective potential
\begin{equation}
  U_{eff}=\mathrm{min}_{n\in P}\frac{e^2_{eff}}{2}|t_{eff}(\sigma)+2\pi i n|^2,
  \end{equation}
where $P$ is the weight lattice of $T$. The Coulomb branch is lifted except at the zeroes of $U_{eff}$.

If $\mathrm{dim}T=1$ there are only Coulomb and Higgs branches. For more general cases such as models with $G=U(1)^k$ ($k>1$) or non-abelian gauge groups there are also mixed branches, and also more complicated phases than we are going to consider there. See \cite{Hori:2016txh} for a recent account in the context of non-abelian GLSMs.
\subsubsection*{One-parameter Calabi-Yau hypersurfaces}
In the following we focus in the case $G=U(1)$, which has already been discussed in Witten's original paper \cite{Witten:1993yc}. The most prominent representative of this class is the GLSM for the quintic in $\mathbb{P}^4$. In the context of the GLSM with B-branes the quintic has been one of the main examples of \cite{Herbst:2008jq}. In connection with the hemisphere partition function the quintic has been used in \cite{Honda:2013uca,Hori:2013ika,Knapp:2016rec}. In this note we focus on the three other GLSMs in this class which describe smooth one-parameter Calabi-Yau hypersurfaces in four-dimensional weighted projective space. The scalars of the chiral multiplets live in 
\begin{equation}
  V=\mathbb{C}(Q_0)\oplus\ldots\oplus\mathbb{C}(Q_5),
\end{equation}
with $Q_0\equiv -N$,  $N=6,8,10$ and $\sum_{i}Q_i=0$ in accordance with the Calabi-Yau condition. We denote the matter fields by $\phi=(p,x_1,\ldots,x_5)$. Concretely the gauge and R-charges of the fields are
\begin{equation}
 \begin{tabular}{c||c|c|c|c|r}
    $N$&$ $&$x_{1 \dots 3}$&$x_{4}$&$x_{5}$& $p$\\
    \hline
   \multirow{2}{*}{6}& $Q:$ & $1$ &$1$ &$2$ &$-6$ \\ 
   &$R:$ &$0$ &$0$ &$0$&$2$ \\
    \hline
   \multirow{2}{*}{8}& $Q:$ & $1$ &$1$ &$4$ &$-8$ \\ 
   &$R:$ &$0$ &$0$ &$0$&$2$ \\
    \hline
   \multirow{2}{*}{10}& $Q:$ & $1$ &$2$ &$5$ &$-10$ \\ 
   &$R:$ &$0$ &$0$ &$0$&$2$
    \end{tabular}
\end{equation}
The superpotential is
\begin{equation}
  W=pG_{N}(x_1,\ldots,x_5),
\end{equation}
where $G_N$ is a generic polynomial of degree $N$ subject to the regularity condition that the equations
\begin{equation}
  \label{regularity}
  \frac{\partial G}{\partial x_{1}} = \dots = \frac{\partial G}{\partial x_{5}} =0
\end{equation}
have no common solutions other than $x_1=\ldots=x_5=0$. There is one field $\sigma$. The classical potential is
\begin{equation}
  U=\frac{1}{2e^{2}} |{D}|^{2} + \sum_{i} |{F_{i}}|^{2} + 2\sigma\bar{\sigma}\sum_{i} Q_{i}^{2} |{\phi_{i}}|^{2}.
\end{equation}
The D-term and F-term equations are
\begin{align}
  D &= -e^2 \left(\sum_{i=1}^{5} Q_{i} |{x_{i}}|^2 - N |p|^{2} -\zeta \right) & F_{1\dots 5} &= p\frac{\partial G_{N}}{\partial x_{1 \dots 5}} &
  F_{p} &= G_{N} \left(x_{1}, \dots x_{5} \right). 
  \end{align}
At $\sigma=0$ we are on the Higgs branch and the low energy theory is determined by the solutions of the D-term and F-term equations. For $\zeta\gg0$ the vacuum manifold is given by
\begin{align}
X^{\zeta\gg0}:\quad \left\{(x_1,\ldots x_5)\in\mathbb{P}(Q_1,\ldots,Q_5)|G_N(x_1,\ldots,x_5)=0\right\} \quad I_{\zeta\gg0}=(x_1,\ldots,x_5)
  \label{geometry}
  \end{align}
This phase is described in terms of a non-linear sigma model with Calabi-Yau target (\ref{geometry}). This phase is referred to as the geometric or large radius phase.

For $\zeta\ll0$ the vacuum is given by 
\begin{equation}
X^{\zeta\ll0}:\quad p=\pm \sqrt{-\frac{\zeta}{N}}, \qquad x_i=0\quad I_{\zeta\ll0}=(p)
\end{equation}
The gauge symmetry is broken to $\mathbb{Z}_N$. Taking into account classical fluctuations around this vacuum, one obtains a potential of the form $W=G_N(x_1,\ldots,x_5)$. This phase describes a Landau-Ginzburg orbifold.

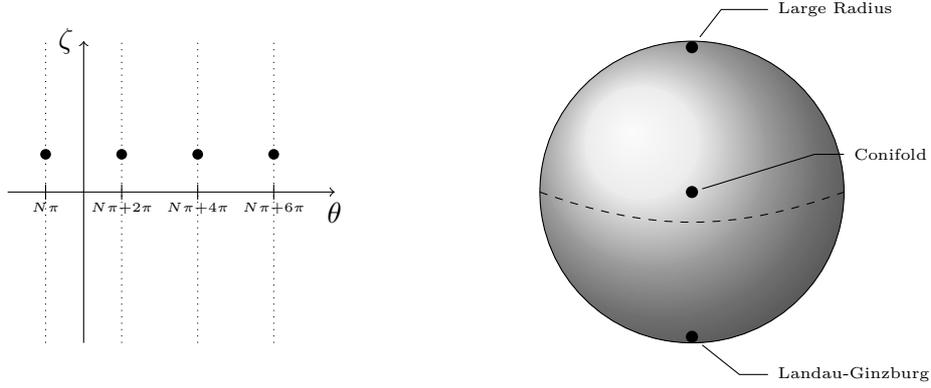
\begin{figure}
\begin{center}

\begin{tikzpicture}
\coordinate (start) at (-1,5);
\draw[<-] (-1,2)--(-1,-2);
\draw[->] (-2,0) -- (2.3,0); 
\node[left] at (-1,2) {$\zeta$};
\node[below] at (2.3,0) {$\theta$};
\def\pos{{-1.5,-0.5,0.5,1.5}};
\def\names{{"$ \scriptstyle N\pi$","$ \scriptstyle N\pi+2\pi$","$\scriptstyle N\pi+4\pi$","$\scriptstyle N\pi+6\pi$"}}
\foreach \i in {0,...,3}
{

\draw[dotted] (\pos[\i],-2)--(\pos[\i],2);
\draw (\pos[\i],-0.1)--(\pos[\i],0.1);
\fill[black] (\pos[\i],0.5) circle [radius=2pt] ;
\node[below] at (\pos[\i],0) {\tiny\pgfmathparse{\names[\i]}\pgfmathresult}; 
}
\coordinate (compact) at (7,0);
\coordinate (calabi) at ($(compact) +(0,1.92)$);
\coordinate (landau) at  ($(compact) -(0,1.92)$);
\filldraw[ball color=gray!20] (compact) circle (2cm);
\fill[black] (calabi) circle [radius=0.08cm];
\fill[black] (compact)  circle [radius=0.08cm];
\fill[black] (landau) circle [radius=0.08cm];
\node[right] (cy) at ($(calabi)+(1,0.5)$) { \tiny Large Radius};
\node[] (cy2) at (calabi){};
\draw (cy)--($(cy)-(1.5,0)$)--(cy2);
\node[right] (lg) at ($(landau)+(1,-0.5)$) { \tiny Landau-Ginzburg};
\node[] (lg2) at (landau){};
\draw (lg)--($(lg)-(1.5,0)$)--(lg2);
\node[right] (con) at ($(compact) +(2,0.5)$) { \tiny Conifold};
\node (cent) at (compact) {};
\draw[-] (con)--($(con)-(1,0)$) -- (cent);
\draw [dashed, bend angle=20, bend right]  ($(compact) -(2,0)$) to ($(compact) +(2,0)$);
\end{tikzpicture}
\end{center}
  \caption{K\"ahler moduli space: FI-theta space (left) and compactified version (right). }\label{fig-modulispace}
\end{figure}

At $\phi=0$ and $\sigma\neq 0$ there is a Coulomb branch. The effective potential is
\begin{align}
  \widetilde{W}_{eff}(\sigma)&= -t \sigma - \sum_{i=0}^{5} (Q_{i} \sigma)\left( \log \left( Q_{i} \sigma \right)-1\right) & \mod &2 \pi i
\end{align}
From this we deduce that the Coulomb branch is localized at
\begin{align}
  e^{-t}&=(-1)^{N}\prod_{i} |{Q_{i}}|^{-Q_{i}}& \rightarrow& &  \zeta&=- \sum_{i} Q_{i} \log |{Q_{i}}|, & \theta&= N \pi + 2 \pi \mathbb{Z}.
\end{align}
In summary, there are three distinct points in the moduli space: the large radius point, the Landau-Ginzburg point and the singular point where the Coulomb branch sits. Under mirror symmetry, the latter gets mapped to the conifold point, or more generally, the discriminant. The situation is depicted in figure \ref{fig-modulispace}. 
For later reference, we also list the topological numbers of the three Calabi-Yaus which we will from now on denote by $X_N:\:\mathbb{P}(Q_1\ldots Q_5)[N]$.
\begin{equation}
  \label{topdata}
  \begin{tabular}{c||c|c|c|c|c}
    $N$&$h^{1,1}(X)$&$h^{2,1}(X)$&$\chi(X)$&$c_2(X)\cdot H$&$H^{3}$\\
    \hline
    $6$ &  $1$ & $103$ & $-204$ & $42$ & $3$ \\
    $8$ & $1$ & $149$ & $-296$ & $44$ & $2$ \\
    $10$ & $1$ & $145$& $-288$ & $34$ & $1$
    \end{tabular}
\end{equation}
Here $h^{i,j}(X)$ are the non-trivial Hodge numbers, $\chi(X)=2(h^{1,1}(X)-h^{2,1}(X))=c_3(X)$ is the Euler number, $H\in H_2(X)$ is the hyperplane class and $c_i(X)$ are the Chern classes.

We can also associate Picard-Fuchs differential operators to each of $X_N$, or rather its mirror $\widetilde{X}_N$
 (see for instance \cite{vanstraten}):
\begin{align}
  N&=6:& \mathcal{L}&=\theta^4-36 z \left(6 \theta+1 \right) \left(3 \theta+1 \right) \left(3 \theta+2 \right) \left(6 \theta+5 \right),\\
   N&=8:& \mathcal{L}&=\theta^4-16 z \left(8 \theta+1 \right) \left(8 \theta+3 \right) \left(8 \theta+5 \right) \left(8 \theta+7 \right),\\
    N&=10:& \mathcal{L}&=\theta^4-80 z \left(10 \theta+1 \right) \left(10 \theta+3 \right) \left(10 \theta+7 \right) \left(10 \theta+9 \right),
\end{align}
where $\theta=z\frac{d}{dz}$ and $z=0$ corresponds to the large complex structure point of $\widetilde{X}_N$. The solutions of the Picard-Fuchs equation $\mathcal{L}\varpi_i(z)=0$ ($i=0,\ldots,3$) are the periods of $\widetilde{X}_N$. For our models, the solutions are given by the following expressions:
\begin{equation}
  \label{periods}
 \varpi_{k} \left(z \right) = \frac{1}{\left(2 \pi i\right)^{k}} \sum\limits_{m=0}^{\infty} \frac{\partial^k}{\partial\epsilon^k} \left[ \frac{\Gamma \left(N(m+\epsilon)+1 \right)}{\Gamma \left(N \epsilon +1 \right)} \prod\limits_{i=1}^{5} \frac{\Gamma \left(Q_{i} \epsilon+1 \right)}{\Gamma \left(Q_{i} (m+ \epsilon) +1 \right)} z^{m+\epsilon} \right] _{\epsilon=0}.
  \end{equation}
\subsection{B-branes and hemisphere partition function}
B-type D-branes in the GLSM are matrix factorizations of the superpotential \cite{Herbst:2008jq,Honda:2013uca,Hori:2013ika}. Define $S=\mathbb{C}[\phi]$ and $M=M^0\oplus M^1$ as a $\mathbb{Z}_2$-graded free S-module -- the Chan-Paton space. A matrix factorization $Q\in\mathrm{End}^1_S(M)$ satisfies
\begin{equation}
  \label{mf}
  Q^2=W\cdot\mathrm{id}_M.
\end{equation}
In addition, $Q$ has to be gauge invariant, and we have to associate R-charge $1$ to it in accordance with the fact the $W$ has R-charge $2$. 
The gauge charges of the D-brane are encoded in $\rho:G\rightarrow GL(M)$ subject to the condition
\begin{equation}
  \label{ggrading}
  \rho(g)^{-1}Q(g\phi)\rho(q)=Q(\phi).
\end{equation}
The R-grading is defined by ${\bf r}_{\ast}:u(1)_R\rightarrow gl(M)$ such that
\begin{equation}
  \label{rgrading}
  \lambda^{{\bf r}_{\ast}}Q(\lambda^R\phi)\lambda^{-{\bf r}_{\ast}}=\lambda Q(\phi).
\end{equation}
This promotes the $\mathbb{Z}_2$-grading to an integer grading. A B-type D-brane in the GLSM is thus described by the quadruple $\mathcal{B}=(M,Q,\rho,{\bf r}_{\ast})$.

A convenient way to describe $M$ is by making use of a Clifford basis. Take a set of $2k$ $2^k\times 2^k$ matrices $\eta_i,\bar{\eta}_i$ satisfying
\begin{equation}
  \label{clifford}
  \{\eta_i,\bar{\eta}_j\}=\delta_{ij}\qquad \{\eta_i,\eta_j\}=\{\bar{\eta}_i,\bar{\eta}_j\}=0
\end{equation}
Define a vacuum $|0\rangle_{q,r}$, where $(q,r)$ is some choice of gauge and $R$-charge. This fixes the overall normalization of (\ref{ggrading}) and (\ref{rgrading}). Declaring the $\bar{\eta}_i$ to be creation operators, the Chan-Paton space $M$ is
\begin{equation}
  M=\bigoplus_{i=0}^k\bigoplus_{m,|m|=k}\prod_{j\in m}\bar{\eta}_j|0\rangle_{q,r},
  \end{equation}
where $m$ is an index set that labels all $\left(\begin{array}{c}k\\j\end{array}\right)$ inequivalent combinations of products of $\bar{\eta}_i$. The matrix factorizations (\ref{mf}) can then also be expressed in terms of the Clifford basis. Note that not all GLSM branes have a description in terms of a Clifford basis. There is no classification of matrix factorizations in any GLSM. There are, however, various constructions of matrix factorizations that lead to particular kinds of D-branes in the various phases. This will be discussed in more detail in section \ref{sec-mfconstruction}.

In the case of abelian GLSMs we can represent any GLSM brane by a complex of Wilson line branes $\mathcal{W}(q)_r$ \cite{Herbst:2008jq}, where $q$ and $r$ denote the gradings encoded in (\ref{ggrading}) and (\ref{rgrading}), respectively. Schematically, it looks like this.
\begin{align}
\mathcal{B}: \quad
\xymatrix{
\cdots \ar@<0.5ex>[r]
  &
 \bigoplus\limits^{L_{j}}_{i=1} \mathcal{W} (q_{j}^{(i)})^{\oplus n_{j}^{(i)}}_{r_{j}^{(i)}}
 \ar@<0.5ex>[r]
 \ar@<0.5ex>[l]
&
\bigoplus\limits^{L_{j+1}}_{i=1} \mathcal{W} (q_{j+1}^{(i)})^{\oplus n_{j+1}^{(i)}}_{r_{j+1}^{(i)}}
\ar@<0.5ex>[r]
 \ar@<0.5ex>[l]
&
\cdots  \ar@<0.5ex>[l]
}
.
\end{align}
The precise form of the maps in the complex is encoded in $Q$. 

Given a GLSM brane $\mathcal{B}$ the hemisphere partition function $Z_{D^2}(\mathcal{B})$ is defined as\footnote{We only consider the Calabi-Yau case here, where the dependence on the radius $r$ of the hemisphere enters trivially.} \cite{Sugishita:2013jca,Honda:2013uca,Hori:2016txh}
\begin{equation}
  \label{zd2}
  Z_{D^2}(\mathcal{B})=C\int_{\gamma}\mathrm{d}^{\mathrm{rk}_G}\sigma\prod_{\alpha>0}\alpha(\sigma)\sinh(\pi\alpha(\sigma))\prod_i\Gamma\left(iQ_i(\sigma)+\frac{R_i}{2}\right)e^{it(\sigma)}f_{\mathcal{B}}(\sigma)
\end{equation}
where
\begin{equation}
  \label{branefactor}
  f_{\mathcal{B}}(\sigma)=\mathrm{tr}_M\left(e^{i\pi{\bf r}_{\ast}}e^{2\pi\rho(\sigma)}\right)
\end{equation}
contains the information about the brane and is referred to as the brane factor. In (\ref{zd2}) $C$ is a normalization constant and the Lagrangian $\gamma\subset\mathfrak{t}_{\mathbb{C}}$ is an integration contour that is a deformation of the real locus $i\mathfrak{t}$ such that the poles of the integrand are avoided and the integral is convergent. Furthermore $\alpha>0$ are the positive roots and $Q_i$ and $R_i$ are the weights of $\rho_V$ and $R$, respectively.
\subsubsection*{One-parameter Calabi-Yau hypersurfaces}
For our examples the hemisphere partition function is
\begin{equation}
  \label{z1par}
        {Z}_{D^2}= C \int_{\gamma} \mathrm{d}\sigma \prod\limits_{i=1}^{5} \Gamma \left(i Q_{i} \left(\sigma\right) \right) \Gamma \left( -iN\left(\sigma\right) + 1 \right) e^{it \sigma} f_{\mathcal{B}} \left(\sigma\right).
\end{equation}
Depending on whether $\zeta\gg0$ or $\zeta\ll0$ we have to close the integration contour in the upper or lower half of the complex plane, respectively. In these two cases the following poles of the integrand contribute to the residue
\begin{align}
  \zeta\gg0&:& \sigma&=i \frac{n}{Q_{i}} & n \in & \mathbb{Z}_{\geq 0} \\
  \zeta\ll0&:& \sigma&=-i \frac{n}{N} & n \in & \mathbb{Z}_{>0} 
\end{align}
The brane factor can be written as
\begin{equation}
  f_{\mathcal{B}}=\sum_{i}e^{i\pi r_i}e^{2\pi\sigma q_i}.
  \end{equation}
In the large radius phases of $U(1)$ GLSMs it has been shown that (\ref{zd2}) computes the quantum corrected central charge of the branes in the phase:
\begin{equation}
  \label{zgeom}
  Z_{D^2}^{\zeta\gg0}(\mathcal{B})=C\sum_{n=0}^{\infty}e^{-nt}\int_X\hat{\Gamma}_X(n)e^{B+\frac{i}{2\pi}\omega}\mathrm{ch}(\mathcal{B}_{LR}).
\end{equation}
Here $\hat{\Gamma}_X$ is the Gamma class, $B+\frac{i}{2\pi}\omega$ is the complexified K\"ahler form on $X^{\zeta\gg0}$ and $\mathrm{ch}(\mathcal{B}_{LR})$ is the Chern character of the brane $\mathcal{B}_{LR}$ in the large radius phase. We will use this information to characterize GLSM branes according to their central charge in the large radius phase. Under the identification $z=e^{-t}$ (\ref{zgeom}) can be expanded in terms of the periods (\ref{periods}) of the mirror Calabi-Yau. Then (\ref{zd2}) for suitably chosen brane factors is nothing but the well-known Mellin-Barnes representation of the mirror periods.
\section{D-brane transport and monodromy in the GLSM}
\label{sec-branes}
In this section we briefly summarize the main results of \cite{Herbst:2008jq}, where the problem of D-brane transport between phases of abelian GLSMs has been solved. Then we give a prescription on how to use this information to compute the monodromy matrices directly from the hemisphere partition function. While our main focus is on $U(1)$ GLSMs, the concepts are more general.
\subsection{GLSM branes vs. branes in phases}
A B-type D-brane in the GLSM is always a matrix factorization of the GLSM superpotential. How this brane is described in the low energy theory of a phase very much depends on the nature of this phase. In the examples we discuss here we are in the lucky situation that the D-branes in both phases are well-understood. In geometric phases B-branes are described in terms of objects of the derived category of coherent sheaves. In the Landau-Ginzburg phase D-branes are matrix factorizations of the Landau-Ginzburg superpotential. In \cite{Herbst:2008jq} one finds a prescription on how to map between GLSM branes and branes in these phases. Since we will do most of our calculations in the GLSM, independently of a specific phase, we refer to the original paper, in particular section 10, for details. For more general phases, such as those arising from non-abelian GLSMs, or more exotic phases like hybrid phases, a good understanding of the D-branes is largely an open issue. 

One important piece of information, which will also be crucial for our discussion, is the fact that the category of GLSM branes is larger than the categories of D-branes associated to the phases. In other words, for a given D-brane in a phase, there are infinitely many GLSM branes which lead to the same low-energy description. Related to this, there are certain GLSM branes $\mathcal{B}_E$ which do not exist in the low-energy theory associated to a phase. We refer to them as ``empty branes''.

In the case of the GLSMs associated to one-parameter Calabi-Yau hypersurfaces the empty branes have been identified. In the large radius phase the empty brane can be written in terms of the $32\times 32$ matrix factorization that can be expressed in terms of the Clifford basis (\ref{clifford}):
\begin{equation}
  \label{lrempty}
  \mathcal{B}_E^{\zeta\gg0}:\quad Q= \sum\limits^{5}_{i=1} \left( x_{i} \eta_{i} + p \frac{1}{d_{i}} \frac{\partial G_N}{\partial x_{i}} \bar{\eta}_{i} \right),
\end{equation}
where $d_{i}$ is the degree of $x_{i}$ in $G_N(x)$. 
Depending on the precise choice of $M$ and $(\rho(g),{\bf r}_{\ast})$ the empty branes can equivalently be written in terms of the following complexes of Wilson line branes for the cases $N=6,8,10$, respectively:
\begin{align}
\xymatrix@1@C=10pt{\mathcal{W}(\tilde{q}_{6})^{\oplus 1}_{r-5}\ar@<2pt>[rr]^-{x}&&\ar@<2pt>[ll]^-{ px^{d-1}}{\begin{array}{c}
 \mathcal{W}(\tilde{q}_{4})^{\oplus 1}_{r-4} \\
 \bigoplus \\
  \mathcal{W}(\tilde{q}_{5})^{\oplus 4}_{r-4}
\end{array}}\ar@<2pt>[rr]^-{x}
&&\ar@<2pt>[ll]^-{ px^{d-1}} {\begin{array}{c}
 \mathcal{W}(\tilde{q}_{3})^{\oplus 4}_{r-3} \\
  \bigoplus \\
  \mathcal{W}(\tilde{q}_{4})^{\oplus 6}_{r-3}
\end{array}}\ar@<2pt>[rr]^-{x}
&&\ar@<2pt>[ll]^-{ px^{d-1}} {\begin{array}{c}
 \mathcal{W}(\tilde{q}_{2})^{\oplus 6}_{r-2} \\
  \bigoplus \\
  \mathcal{W}(\tilde{q}_{3})^{\oplus 4}_{r-2}
\end{array}}\ar@<2pt>[rr]^-{x}
&&\ar@<2pt>[ll]^-{ px^{d-1}} {\begin{array}{c}
 \mathcal{W}(\tilde{q}_{1})^{\oplus 4}_{r-1} \\
  \bigoplus \\
  \mathcal{W}(\tilde{q}_{2})^{\oplus 1}_{r-1}
\end{array}}\ar@<2pt>[rr]^-{x}
&&
\ar@<2pt>[ll]^-{ px^{d-1}}\mathcal{W}(q)^{\oplus 1}_{r}
}
\end{align}
\begin{align}
\xymatrix@1@C=10pt{\mathcal{W}(\tilde{q}_{8})^{\oplus 1}_{r-5}\ar@<2pt>[rr]^-{x}\ar@_{<-}&&\ar@<2pt>[ll]^-{ px^{d-1}}{\begin{array}{c}
 \mathcal{W}(\tilde{q}_{7})^{\oplus 4}_{r-4} \\
  \bigoplus \\
  \mathcal{W}(\tilde{q}_{4})^{\oplus 1}_{r-4}
\end{array}}\ar@<2pt>[rr]^-{x}
&&\ar@<2pt>[ll]^-{ px^{d-1}} {\begin{array}{c}
 \mathcal{W}(\tilde{q}_{6})^{\oplus 6}_{r-3} \\
  \bigoplus \\
  \mathcal{W}(\tilde{q}_{3})^{\oplus 4}_{r-3}
\end{array}}\ar@<2pt>[rr]^-{x}
&& \ar@<2pt>[ll]^-{ px^{d-1}}{\begin{array}{c}
 \mathcal{W}(\tilde{q}_{5})^{\oplus 4}_{r-2} \\
  \bigoplus \\
  \mathcal{W}(\tilde{q}_{2})^{\oplus 6}_{r-2}
\end{array}}\ar@<2pt>[rr]^-{x}
&&\ar@<2pt>[ll]^-{ px^{d-1}} {\begin{array}{c}
 \mathcal{W}(\tilde{q}_{4})^{\oplus 1}_{r-1} \\
  \bigoplus \\
  \mathcal{W}(\tilde{q}_{1})^{\oplus 4}_{r-1}
\end{array}}\ar@<2pt>[rr]^-{x}
&&
\ar@<2pt>[ll]^-{ px^{d-1}}\mathcal{W}(q)^{\oplus 1}_{r}
}
\end{align}
\begin{align}
\xymatrix@1@C=10pt{\mathcal{W}(\tilde{q}_{10})^{\oplus 1}_{r-5}\ar@<2pt>[rr]^-{x} &&\ar@<2pt>[ll]^-{ px^{d-1}}{\begin{array}{c}
 \mathcal{W}(\tilde{q}_{9})^{\oplus 3}_{1} \\
  \bigoplus \\
  \mathcal{W}(\tilde{q}_{8})^{\oplus 1}_{r-4} \\
   \bigoplus \\
  \mathcal{W}(\tilde{q}_{5})^{\oplus 1}_{r-4}
\end{array}}\ar@<2pt>[rr]^-{x}
&& \ar@<2pt>[ll]^-{ px^{d-1}}{\begin{array}{c}
 \mathcal{W}(\tilde{q}_{8})^{\oplus 3}_{r-3} \\
  \bigoplus \\
  \mathcal{W}(\tilde{q}_{7})^{\oplus 3}_{r-3} \\
   \bigoplus \\
  \mathcal{W}(\tilde{q}_{4})^{\oplus 3}_{r-3} \\
   \bigoplus \\
  \mathcal{W}(\tilde{q}_{3})^{\oplus 1}_{r-3}
\end{array}}\ar@<2pt>[rr]^-{x}
&& \ar@<2pt>[ll]^-{ px^{d-1}}{\begin{array}{c}
 \mathcal{W}(\tilde{q}_{7})^{\oplus 1}_{r-2} \\
  \bigoplus \\
  \mathcal{W}(\tilde{q}_{6})^{\oplus 3}_{r-2} \\
   \bigoplus \\
  \mathcal{W}(\tilde{q}_{3})^{\oplus 3}_{r-2} \\
   \bigoplus \\
  \mathcal{W}(\tilde{q}_{2})^{\oplus 3}_{r-2}
\end{array}}\ar@<2pt>[rr]^-{x}
&&\ar@<2pt>[ll]^-{ px^{d-1}} {\begin{array}{c}
 \mathcal{W}(\tilde{q}_{5})^{\oplus 1}_{r-1} \\
  \bigoplus \\
  \mathcal{W}(\tilde{q}_{2})^{\oplus 1}_{r-1} \\
   \bigoplus \\
  \mathcal{W}(\tilde{q}_{1})^{\oplus 3}_{r-1}
\end{array}}\ar@<2pt>[rr]^-{x}
&&
\ar@<2pt>[ll]^-{ px^{d-1}}\mathcal{W}(q)^{\oplus 1}_{r}.
}
\end{align}
Here the integers $q,r$ are fixed by the overall normalization of $(\rho(g),{\bf r}_{\ast})$ and $\tilde{q}_{n}=q-n$.  Furthermore we used the short-hand notation $x=\sum_ix_i\eta_i$, $px^{d-1}=\sum_i p\frac{1}{d_i} \frac{\partial G_N}{\partial x_i}\bar{\eta}_i$. This gives a family of branes. Using (\ref{branefactor}) one finds the following brane factors
\begin{align}
\begin{aligned}
N&=6: & f_{\mathcal{B}_E^{\zeta\gg0}}&=\left(e^{2 \pi  \sigma }-1\right)^5 \left(e^{2 \pi  \sigma }+1\right) e^{\pi  (2 (q-6) \sigma +i r)} \\
 N&=8: & f_{\mathcal{B}_E^{\zeta\gg0}}&= \left(e^{2 \pi  \sigma }-1\right)^5 \left(e^{2 \pi  \sigma }+e^{4 \pi  \sigma }+e^{6 \pi  \sigma }+1\right) e^{\pi  (2 (q-8) \sigma +i r)}\\
 N&=10: &  f_{\mathcal{B}_E^{\zeta\gg0}}&=\left(e^{2 \pi  \sigma }-1\right)^5 \left(2 e^{2 \pi  \sigma }+2 e^{4 \pi  \sigma }+2 e^{6 \pi  \sigma }+2 e^{8 \pi  \sigma }+e^{10 \pi  \sigma }+1\right) e^{\pi  (2 (q-10) \sigma +i r)}.
\end{aligned}
\end{align}
Inserting this into (\ref{z1par}) and closing the contour in the upper half plane all the poles inside the contour are canceled by the brane factor and therefore $Z_{D^2}^{\zeta\gg0}(\mathcal{B}_E^{\zeta\gg0})=0$. Note however, that we get a non-zero result if we close the integration contour the other way.

Conversely, there is also a family of empty branes associated to the Landau-Ginzburg phase given by the $2\times 2$ matrix factorization
\begin{equation}
  \label{lgempty}
  \mathcal{B}_E^{\zeta\ll0}:\quad Q=\begin{pmatrix}
0 & G_N\left(x\right) \\
p & 0
\end{pmatrix}.
\end{equation}
In terms of complexes of Wilson line branes this can be expressed as follows
\begin{equation}
\xymatrix@1@C=20pt{
\mathcal{W} \left(q-N\right)_{r-1} \ar@<0.5ex>[r]^-{G}
&
\mathcal{W}\left(q \right)_{r} \ar@<0.5ex>[l]^-{p}
}.
\end{equation}
The brane factors are
\begin{equation}
  f_{\mathcal{B}_E^{\zeta\ll0}}=e^{i \pi r}e^{2\pi q\sigma} \left(1- e^{-2 \pi N \sigma} \right),
\end{equation}
and one can easily convince oneself that $Z_{D^2}^{\zeta\ll0}(\mathcal{B}_E^{\zeta\ll0})=0$. One observes that ideals defined by the deleted sets $I_{\zeta}$ defined in (\ref{quotient}) are encoded in the matrix factorizations of the empty branes.

\begin{figure}
\centering
\begin{tikzpicture}
\node  at (-4,4) {UV: GLSM};
\node (B1)at (-1.5,4) {$\mathcal{B}_{1}$};
\node (B2) at (-0,4) {$\mathcal{B}_{2}$};
\node (ELG) at (1.5,4) {$\mathcal{B}^{\zeta \ll 0}_{E}$};
\node (ECY) at (3,4) {$\mathcal{B}^{\zeta \gg 0}_{E}$};

\node[rounded corners=8pt, fill=gray!50] at (-1.7,0.2) {CY};
\node (B1CY) at (-0.5,0) {$\mathcal{O}_{X}$};
\node (B2CY) at (-2,0.8) {$\widetilde{\mathcal{O}}$};
\node (B3CY) at (-1.5,-0.8) {$\widetilde{\widetilde{\mathcal{O}}}$};
\node[] (NOB2) at (-0.5,1.5){x};
\node[rounded corners=8pt, fill=gray!50] at (1,-0.5) {LG};
\node (B1LG) at (3.5,0) {$\mathcal{MF}$};
\node (B2LG) at (2,0.2) {$\widetilde{\mathcal{MF}}$};
 \node[] (NOB1) at (2.7,1.5){x};
\node at (-4,0) {IR:};
\draw[thick] (1,1.5)--(0.5,0)--(-1,-1.5);

\draw (ELG)--(B1CY);
\draw[dotted] (ELG) -- (NOB1);
\draw[dotted] (ECY) -- (NOB2);
\draw (ECY) -- (B1LG);
\draw(B1) -- (B2LG);
\draw(B1) -- (B2CY);
\draw(B2) -- (B2LG);
\draw(B2) -- (B3CY);
\end{tikzpicture}
 \caption{Different GLSM branes can map to the same objects in a phase.}\label{fig-lowenergy}
\end{figure}
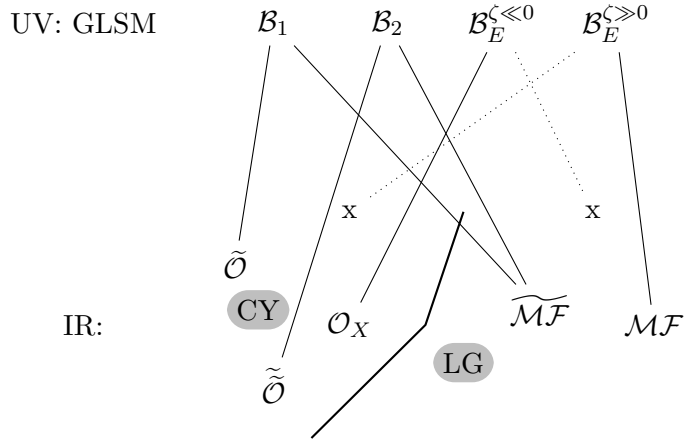

The empty branes also give some intuition why the category of GLSM branes is so much larger than the category of branes associated to a phase. Since we are dealing with triangulated categories, one can form bound states of branes using the cone construction. One can in particular form a bound state between some GLSM brane $\mathcal{B}$ and an empty brane $\mathcal{B}_E^{A}$ associated to some phase $A$. At the level of the GLSM the bound state is different from $\mathcal{B}$. However, when going to phase $A$ (and only phase $A$) both $\mathcal{B}$ and its bound state with $\mathcal{B}_E^{A}$ will map to the same object in the D-brane category associated to phase $A$. In this way, an infinite number of GLSM branes can map to the same object in a phase. The situation for our models is depicted in figure \ref{fig-lowenergy}

Given some generic brane in the GLSM, it will typically reduce to some low energy brane in each of the phases. However, in general such a situation does not describe a well-defined analytic continuation of a low-energy brane from one phase to another. The extreme case are the empty branes $\mathcal{B}_E$ which typically standard branes in one phase and ``nothing'' in another. In our examples the $\mathcal{B}_E^{\zeta\gg0}$ reduces to a matrix factorization describing a Recknagel-Schomerus boundary state with label $L=(0,0,0,0,0)$ in the Landau-Ginzburg phase, while $\mathcal{B}_E^{\zeta\ll0}$ becomes the structure sheaf $\mathcal{O}_X$ in the geometric phase \cite{Herbst:2008jq}. Both of these branes are known to have analytic continuations to the other phases that do not give zero. This implies that the GLSM branes like $\mathcal{B}_E^{\zeta\ll0}$ and $\mathcal{B}_E^{\zeta\gg0}$ are not suitable choices of GLSM branes in the context of analytic continuation between phases. This issue is resolved by the grade-restriction rule \cite{Herbst:2008jq}, which we will recall now.
\subsection{D-brane transport and grade restriction}
\label{sec-grr}
Let us explain how to transport branes between phases. Due to presence of a singular locus at the phase boundary the result of analytic continuation will depend on the choice of path. As depicted in figure \ref{fig-modulispace}, the parameter space is split into chambers of width $2\pi$. Paths from one phase to another, parametrized by $\theta$-angles within two adjacent chambers, differ by a monodromy around the singular point.  

Depending on the choice of path only certain GLSM branes can be transported from one phase to another in a well-defined way. These have to satisfy the grade restriction rule, which states that, depending on the choice of path/$\theta$-angle only GLSM branes with particular gauge charges $q$, as encoded in (\ref{ggrading}), can be safely transported along a given path. For the GLSMs associated to one-parameter Calabi-Yau hypersurfaces the rule is 
\begin{equation}
\label{chargegrr}
w_k:\qquad   \theta\in(-\left(N+2 k \right)\pi,-\left(N+2k \right)\pi+2\pi):\qquad q\in\{k,k+1,\ldots, k+N-1\}\; k\in\mathbb{Z}.
\end{equation}
For every path in the parameter space there is a ``window'' of allowed charges. Since there is a one-to-one correspondence between inequivalent families of paths and allowed charges we will use the terms ``path'' and ``window'' interchangeably for the datum $w_k$. In \cite{Hori:2013ika} it has been shown that this grade restriction rule can be obtained by analyzing the asymptotic behavior of the hemisphere partition function.

At the level of D-brane categories the grade restriction rule is to be understood as follows. For every choice of path/charge window $w_k$, there is a subcategory $D_{w_k}\subset D_{GLSM}$, often referred to as window category, containing those branes which satisfy the grade restriction rule. Each of the categories $D_{w_k}$ is equivalent to the D-brane categories in each of the phases, which then establishes equivalences between the D-brane categories of the respective phases. In the case of Calabi-Yau GLSMs with $G=U(1)$ this produces the equivalence between the category of Landau-Ginzburg branes $D_{LG}$ and the derived category of coherent sheaves $D_{CY}$ \cite{MR2641200}. 

A general GLSM brane typically does not satisfy the grade restriction rule. However, given a GLSM brane $\mathcal{B}$ that descends to a particular low energy brane $\mathcal{B}^A$ in a given phase $A$, we can replace $\mathcal{B}$ by a different GLSM brane $\mathcal{B}_{w_k}$ which also reduces to the brane $\mathcal{B}^A$ in phase $A$, but satisfies the grade restriction rule associated to a given window $w_k$. In contrast to $\mathcal{B}$, $\mathcal{B}_{w_k}$ is well-defined in the low-energy theory beyond the phase boundary. If the path $w_k$ connects phase $A$ with another phase $B$, then the low-energy brane $\mathcal{B}^B$, obtained by reducing $\mathcal{B}_{w_k}$ to phase $B$, is the analytic continuation of $\mathcal{B}^A$ from phase $A$ to phase $B$ along $w_k$. Choosing a different path $w_{k'}$ with different grade restriction rule from $A$ to $B$ leads to a different GLSM lift $\mathcal{B}_{w_k'}$ and a different analytic continuation $\mathcal{B}^{'B}$ to phase $B$ that is related to $\mathcal{B}^B$ by a monodromy around the phase boundary between $A$ and $B$. This is illustrated in figure \ref{fig:mapping}.
\newcommand{\yslant}{0.5}
\newcommand{\xslant}{-0.8}
\begin{figure}
\centering
\begin{tikzpicture}[scale=0.8,decoration={
markings,
mark=between positions 0 and 1 step 0.10 with {\arrow{stealth}},
}
]
	\begin{scope}[
		yshift=-100,
		every node/.append style={yslant=\yslant,xslant=\xslant},
		yslant=\yslant,xslant=\xslant
	] 
		\draw[black, dashed, thin] (0,0) rectangle (4,4);
		\node[] at (0,4.2) {Low Energy};
		 \draw[dotted,thick] (0,2)--(4,2);

		 \node[rounded corners=8pt, fill=gray!50] at (0.5,3.5) {B};
		 \node[rounded corners=8pt, fill=gray!50] at (3,1) {A};
		 \node[] (BA) at (1,1) {$\mathcal{B}^{A}$};
		  \node[] (BB) at (3,2.5) {$\mathcal{B}^{B}$};
		   \node[] (BB2) at (1.5,3) {$\mathcal{B}^{'B}$};

	\end{scope}

	\begin{scope}[
		yshift=0,
		every node/.append style={yslant=\yslant,xslant=\xslant},
		yslant=\yslant,xslant=\xslant
	]
		\fill[white,fill opacity=.55] (0,0) rectangle (4,4); 
		\draw[black, dashed, thin] (0,0) rectangle (4,4); 
		\node[] at (0,4.2) {GLSM};
		
		  \node[] (BAG) at (2,2) {$\mathcal{B}$};
		   \node[] (BWK) at (3,2) {$\mathcal{B}_{w_{k}}$};
		    \node[] (BWK2) at (1,2) {$\mathcal{B}_{w_{k'}}$};
	\end{scope} 
	
	\draw[->](BA) to (BAG);
	\draw[red,thick,->,postaction={decorate}]  (BB) -- (BWK) --(BA) -- (BWK2) --(BB2); 
\end{tikzpicture}\caption{Analytic continuation of a brane in phase $A$ to phase $B$ via different windows.}\label{fig:mapping}
\end{figure}
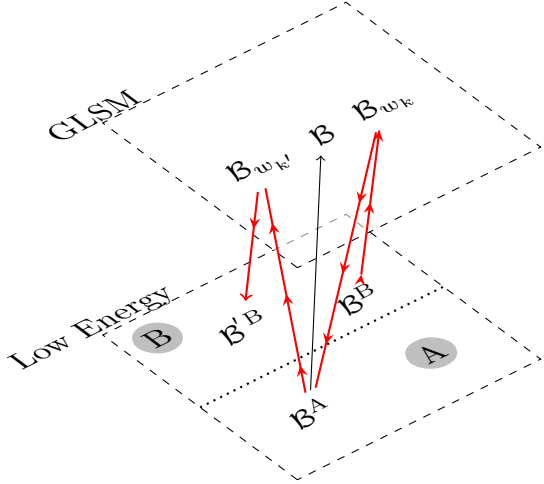

To obtain the grade restricted brane $\mathcal{B}_{w_k}$ from the non-grade restricted one $\mathcal{B}$, one uses the empty brane $\mathcal{B}_{E}^A$ associated to the phase $A$, where $\mathcal{B}$ and $\mathcal{B}_{w_k}$ should reduce to the same low-energy brane $\mathcal{B}^A$. Using the cone construction one forms a bound state $\mathcal{B}'=\mathrm{Cone}(\Psi:\mathcal{B}\rightarrow\mathcal{B}_E^A)$, where the tachyon $\Psi\in\mathcal{H}^1(\mathcal{B},\mathcal{B_E^A})$ is a $\mathbb{Z}_2$-odd boundary changing state defined by the condition
\begin{equation}
  Q_{E}^A\Psi+\Psi Q=0\qquad \Psi\neq Q_E^A\Phi-\Phi Q,
\end{equation}
where $Q$ and $Q_{A}^A$ are the GLSM matrix factorizations associated to $\mathcal{B}$ and $\mathcal{B}_E^A$, respectively, and $\Phi$ is $\mathbb{Z}_2$-even with polynomial entries. To grade-restrict one requires a particular $\Psi$ such that the bound state contains trivial brane-antibrane pairs characterized by
\begin{equation}
  \xymatrix{\mathcal{W}(q)_r\ar@<2pt>[r]^{1} & \ar@<2pt>[l]^{W}\mathcal{W}(q)_{r+1} }\quad\leftrightarrow\quad Q=\left(\begin{array}{cc}0&1\\W&0\end{array}\right)
  \end{equation}
between those Wilson line branes in $\mathcal{B}$ that violate the grade restriction rule. Such trivial brane-antibrane pairs can be removed from the complex describing the bound state $\mathcal{B}'$. Schematically, the procedure works as follows. Assume for simplicity that $\mathcal{B}$ is described as a complex of Wilson line branes, where only one Wilson line brane $\mathcal{W}(q^{\ast})_r$ that sits at the right-most entry of a complex violates the grade restriction rule, i.e.
\begin{equation}
 \mathcal{B}:\quad  \xymatrix{\ldots \ar@<2pt>[r]^{a_2} & \ar@<2pt>[l]^{b_2}\bigoplus_{k_1}\mathcal{W}(q_{k_1})_{r_1}\ar@<2pt>[r]^{a_1} & \ar@<2pt>[l]^{b_1}\bigoplus_{k_0}  \mathcal{W}(q_{k_0})_{r_0}\ar@<2pt>[r]^{a_0} & \ar@<2pt>[l]^{b_0}\mathcal{W}(q^{\ast})_{r} },
  \end{equation}
with $\{q_{k_0},q_{k_1},\ldots\}$ in an allowed charge window. Then, choose $\mathcal{B}^A_E$ such that $q^{\ast}$ appears with multiplicity one, and R-charge $r-1$:
\begin{equation}
  \mathcal{B}^A_E:\quad  \xymatrix{\ldots \ar@<2pt>[r]^{\tilde{a}_2} & \ar@<2pt>[l]^{\tilde{b}_2}\bigoplus_{\tilde{k}_1}\mathcal{W}(\tilde{q}_{\tilde{k}_1})_{\tilde{r}_1}\ar@<2pt>[r]^{\tilde{a}_1} & \ar@<2pt>[l]^{\tilde{b}_1}\bigoplus_{\tilde{k}_0}  \mathcal{W}(\tilde{q}_{\tilde{k}_0})_{\tilde{r}_0}\ar@<2pt>[r]^{\tilde{a}_0} & \ar@<2pt>[l]^{\tilde{b}_0}\mathcal{W}(q^{\ast})_{r-1}},
\end{equation}
where we further assume that  $\{\tilde{q}_{\tilde{k}_0},\tilde{q}_{\tilde{k}_1},\ldots\}$ are in the desired charge window. Then the bound state $\mathcal{B}'$ must have the following form:
\begin{equation}
  \mathcal{B}':\quad \xymatrix{&\ldots \ar@<2pt>[r]^{a_2} & \ar@<2pt>[l]^{b_2}\bigoplus_{k_1}\mathcal{W}(q_{k_1})_{r_1}\ar@<2pt>[r]^{a_1} & \ar@<2pt>[l]^{b_1}\bigoplus_{k_0}  \mathcal{W}(q_{k_0})_{r_0}\ar@<2pt>[r]^{a_0} & \ar@<2pt>[l]^{b_0}{\color{purple}\underline{\mathcal{W}(q^{\ast})_{r}}} \\
\ldots \ar@<2pt>[r]^{\tilde{a}_2} & \ar@<2pt>[l]^{\tilde{b}_2}\bigoplus_{\tilde{k}_1}\mathcal{W}(\tilde{q}_{\tilde{k}_1})_{\tilde{r}_1}\ar@<2pt>[r]^{\tilde{a}_1} \ar@<2pt>[ru]_{\varphi_1}  & \ar@<2pt>[l]^{\tilde{b}_1}\bigoplus_{\tilde{k}_0}  \mathcal{W}(\tilde{q}_{\tilde{k}_0})_{\tilde{r}_0}\ar@<2pt>[r]^{\tilde{a}_0}\ar@<2pt>[ru]_{\varphi_0} & \ar@<2pt>[l]^{\tilde{b}_0}{\color{purple}\underline{\mathcal{W}(q^{\ast})_{r-1}}}\ar@<2pt>[ru]_{1} & }
  \end{equation}
Here the diagonal maps encode the information about the tachyon $\Psi$. The pair of Wilson line branes (in color and underlined) connected by the identity map is a trivial brane-antibrane pair and can be eliminated. Thus, $\mathcal{B}'$ is grade restricted. $\mathcal{B}$ and $\mathcal{B}'$ have the same low-energy behavior in phase $A$ where $\mathcal{B}^A_E$ is empty. For branes where more than one Wilson line brane violates the grade restriction rule, this procedure has to be iterated. 

Only the charges and multiplicities of the Wilson line branes enter the brane factor (\ref{branefactor}) of the hemisphere partition function. The precise form of the maps encoded in $\Psi$ is not required. {\em Assuming} that a tachyon $\Psi$ with the desired properties can always be found, one can construct the brane factor of a grade restricted brane by simply placing the complexes of the branes involved in the correct relative positions. This pragmatic approach was taken in \cite{Knapp:2016rec}, and we will also apply it here. 
\subsection{Monodromy and hemisphere partition function}
\label{sec-zd2monodromy}
Having full control over the paths along which we transport the D-branes, we can compute monodromies of D-branes around the distinguished loci in the moduli space. A discussion on how to do monodromies using the GLSM has already been given in section 10.5 of \cite{Herbst:2008jq}. Combining these methods with the hemisphere partition function makes is possible to obtain the monodromy matrices associated to a Calabi-Yau directly from the brane factors. This approach was applied to compute large radius and conifold monodromies on the quintic in \cite{Knapp:2016rec}. Here we give a more general and detailed description.

The idea is to take a basis $\mathcal{B}_i$ of GLSM branes, to compute their monodromy images $\widetilde{\mathcal{B}}'_i$ and to read off the components $M_{ij}$ of the monodromy matrix $M$ from the relation between the respective brane factors:
\begin{equation}
f_{\widetilde{\mathcal{B}}'_i}=\sum_jM_{ij}f_{\mathcal{B}_j}.
\end{equation}
The discussion can be done entirely in the GLSM and at the level of the brane factors. Once we have chosen a basis, it is not necessary to evaluate the hemisphere partition function in a particular phase.

The first step is to pick a suitable set of GLSM branes. When we discuss concrete examples in section \ref{sec-monodromy} we will choose the basis such that the branes in the large radius phase are simple examples of (D0,D2,D4,D6)-branes. One could however, equally well construct a set of GLSM branes that reduces to a nice basis in any phase. We identify our basis of D-branes by explicitly constructing matrix factorizations in the GLSM. To confirm that these are a good basis, we evaluate the hemisphere partition function in the large volume phase to see whether we obtain a linearly independent set of brane charges. In this way we can avoid the often tedious procedure of explicitly mapping GLSM branes to branes in the low-energy phase.

In order to discuss the monodromy we have to fix a reference point. At the level of the GLSM this is done by choosing a reference path $w_k$ in the moduli space, or equivalently, a particular set of allowed charges via the grade restriction rule. This not only fixes a reference point for the monodromy for one particular phase, but rather gives us a particular family of reference points all over the moduli space. If any of the GLSM branes of our basis is not grade-restricted, we have to grade restrict them first to $w_k$ by using the methods described in section \ref{sec-grr}. To grade restrict, we have to use the empty brane associated to the phase with respect to which we have chosen our basis.

Once we have picked a basis of branes and a window $w_k$ we can perform the monodromy. The procedure depends on whether the path along which we transport the D-brane crosses a phase boundary or not. In our examples the former applies to the large radius and Landau-Ginzburg points, the latter applies to the monodromy around the singular point.
\subsubsection*{Monodromies inside a phase}
When the path of the D-brane does not cross a phase boundary, the monodromy operation amounts to a $2\pi$-shift of the theta angle. In the hemisphere partition function the theta angle appears in the the term $e^{it(\sigma)}$. In the one-parameter case we have
\begin{equation}
  e^{it\sigma}=e^{i(\zeta-i\theta)\sigma}\;\stackrel{\theta\rightarrow\theta+2\pi}{\longrightarrow}\;e^{i(\zeta-i(\theta+2\pi))\sigma}=e^{it\sigma}e^{2\pi\sigma}.
\end{equation}
The shift by $e^{2\pi\sigma}$ can be absorbed into the brane factor $f_{\mathcal{B}}(\sigma)$:
\begin{equation}
  \theta\rightarrow\theta+2\pi:\qquad f_{\mathcal{B}}(\sigma)\longrightarrow f_{\mathcal{B}'}(\sigma)=e^{2\pi\sigma}f_{\mathcal{B}}(\sigma).
\end{equation}
Thinking of $\mathcal{B}$ as a particular complex of Wilson line branes, its monodromy image $\mathcal{B}'$ is a complex of Wilson line branes where all the gauge charges are shifted by $+1$, i.e. the whole complex gets tensored by $\mathcal{W}(1)$. Typically, $\mathcal{B}'$ will no longer satisfy the grade restriction rule associated to the reference path. If $\mathcal{B}$ was grade restricted with respect to the charge window $w_k$, $\mathcal{B}'$ will be grade restricted with respect to the adjacent window $w_{k+1}$. Since $w_{k+1}$ is associated a path to the left of the path associated to $w_k$, a shift $\theta\rightarrow\theta+2\pi$ in the FI-$\theta$ parameter is equivalent to moving the brane to the left in the $\theta$-direction. In this case one cannot decide whether the brane moves around the distinguished point in clockwise or counter-clockwise direction. However, if we make the same shift for two adjacent phases the relative orientations of the contours are opposite. 

To compare with the original set of branes we have to apply the grade restriction procedure to replace $\mathcal{B}'$ by a GLSM brane $\widetilde{\mathcal{B}}'$ that is grade restricted to the window $w_k$ we started with. We have to use the empty branes associated to the phase under consideration. The brane factor $f_{\widetilde{\mathcal{B}}'}$ can then be expressed in terms of the brane factors of the original basis of GLSM branes. The coefficients encode the information about the monodromy.

In summary, we have the following recipe for computing the monodromy around a point inside a phase $A$:
\begin{enumerate}
\item Pick a basis $\mathcal{B}_i$ of GLSM branes that is grade-restricted with respect to a particular path $w_k$. 
\item Compute the monodromy images $\mathcal{B}_i'$ of each basis element by shifting every gauge charge $q$ of $\mathcal{B}_i$ by $+1$. The resulting brane fits into the window $w_{k+1}$.
\item Grade restrict $\mathcal{B}_i'$ to $w_k$ by using the empty branes $\mathcal{B}_{E}^A$ associated to phase $A$. This yields a set of branes $\widetilde{\mathcal{B}}_i'$. 
  \item Read off the coefficients $M_{ij}$ of the monodromy matrix $M$  from the relation $f_{\widetilde{\mathcal{B}}'_i}=\sum_jM_{ij}f_{\mathcal{B}_j}$ between the brane factors.
\end{enumerate}
There is also another way to compute the monodromy matrices in this situation. Instead of comparing the brane factors, we could also choose the basis $\mathcal{B}_i$ with respect to phase $A$ and evaluate the hemisphere partition function in $A$. Then we can read off the monodromy matrix from
\begin{equation}
  Z_{D^2}^A(\mathcal{B}_i')=\sum_{ij}M_{ij}Z_{D^2}^A(\mathcal{B}_j).
\end{equation}
In this case neither $\mathcal{B}_i$ nor the image branes $\mathcal{B}_i'$ need to be grade restricted. Since the evaluation of the contour integrals in a phase can be tedious, this approach is not necessarily simpler.
\subsubsection*{Monodromy across phase boundaries}
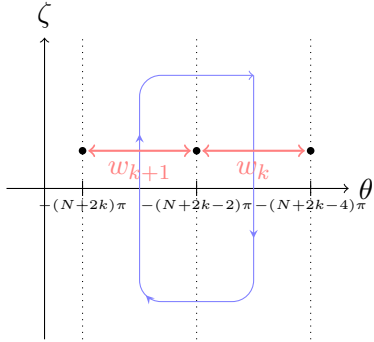
\begin{figure}
\centering
\begin{tikzpicture}[decoration={
markings,
mark=at position 0.25 with {\arrow{stealth}},
mark=at  position 0.50 with {\arrow{stealth}},
mark=at  position 0.75 with {\arrow{stealth}},
}]
    \draw[thin, ->] (-1,0) -- (3.5,0) node[anchor=west]{$\theta$};
    \draw[thin, ->] (-0.5,-2) -- (-0.5,2) node[anchor=south]{$\zeta$};
    \def\names{{"$ \scriptstyle -(N+2k)\pi$","$\scriptstyle -(N+2k-2)\pi$","$\scriptstyle -(N+2k-4)\pi$"}};
  \foreach \x in {0,...,2}{
   \draw[dotted] (\x*1.5,-2) -- (\x*1.5,2);
   \draw (\x*1.5,0.1) -- (\x*1.5,-0.1);
	\node[below] at (\x*1.5,0) {\tiny\pgfmathparse{\names[\x]}\pgfmathresult}; 
	 \fill[black] (\x*1.5,0.5) circle (0.05);
  	}
  	
  	\draw[thin,blue!50,rounded corners=8pt,->,postaction={decorate}] (1.5*1.5,1.5)--(1.5*1.5,-1.5)--(0.5*1.5,-1.5)--(0.5*1.5,1.5)--(1.5*1.5,1.5);
  	\draw[thick, red!50,<->] (0.05*1.5,0.5)--(0.5*1.5,0.5) node[below] {$w_{k+1}$}--(0.95*1.5,0.5) ;
  	\draw[thick, red!50,<->] (1.05*1.5,0.5)--(1.5*1.5,0.5) node[below] {$w_{k}$}--(1.95*1.5,0.5) ;

\end{tikzpicture}
  \caption{Monodromy around a point at a phase boundary.}\label{fig-conifoldmonodromy}
\end{figure}

If we compute a monodromy around a singular locus at a phase boundary, we are forced to leave a particular phase. Then the monodromy cannot simply be encoded in a shift of the theta angle in the integrand of $Z_{D^2}$. An example of a conifold monodromy in our one-parameter models is depicted in figure \ref{fig-conifoldmonodromy}. 
The theta angle shift required for encircling the singular point is now achieved by grade restricting the branes according to two adjacent paths in the parameter space \cite{Herbst:2008jq}.

As in the other case we start off with a grade restricted brane $\mathcal{B}$ that is well-defined along the full path across $w_k$ we have chosen. To encircle a singular point we grade restrict with respect to an adjacent path $w_{k\pm1}$, depending on whether we want to move the brane to the left or to the right. Which empty brane we use to grade-restrict determines the reference point. For example, if we start in the large radius phase we have to perform the window shift using the empty brane of the Landau-Ginzburg phase in order to make sure that the contour has its turning point in the $\zeta<0$ phase. We get a brane $\mathcal{B}'$ which is now defined along the full path $w_{k\pm1}$. To complete the calculation, we have to grade restrict back to $w_k$ using the empty brane of the phase we have put our reference point in. This yields a brane $\widetilde{\mathcal{B}}'$ whose brane factor can then be expressed in terms of the brane factors of the basis of branes. From this we can read off the entries in the monodromy matrix.

In summary, we get the following recipe for a monodromy around a singularity at a phase boundary between phases $A$ and $B$ that sits between the two windows $w_k$ and $w_{k+1}$.
\begin{enumerate}
\item Pick a basis $\mathcal{B}_i$ that is grade-restricted with respect to $w_k$. Assume without loss of generality that the reference point is in phase $A$.
\item  Grade restrict $\mathcal{B}_i$ to $w_{k+1}$ using the empty branes $\mathcal{B}_{E}^B$ associated to phase $B$. This yields a set of branes $\mathcal{B}_i'$.
\item Grade restrict $\mathcal{B}_i'$ to $w_{k}$ using the empty branes $\mathcal{B}_{E}^A$ associated to phase $A$. This yields a set of branes $\widetilde{\mathcal{B}}_i'$.
\item Read off the coefficients $M_{ij}$ of the monodromy matrix $M$  from the relation $f_{\widetilde{\mathcal{B}}'_i}=\sum_jM_{ij}f_{\mathcal{B}_j}$ between the brane factors.
\end{enumerate}
If we compared the hemisphere partition functions $Z_{D^2}^A$ in phase $A$ instead of the brane factors, we could skip the third step. If the reference point were in phase $B$, we would have to use $\mathcal{B}_{E}^A$ in step $2$ and  $\mathcal{B}_{E}^B$ in step $3$. Note that in this case the direction of the contour changes from clockwise to counter-clockwise or vice versa.  
\section{Constructions of GLSM branes}
\label{sec-onepar}
In this section we discuss explicit constructions of matrix factorizations, focusing on the GLSMs associated to Calabi-Yau hypersurfaces in weighted projective space. This is inspired by D-brane constructions in Landau-Ginzburg and Calabi-Yau phases. While there is no classification of matrix factorizations in GLSMs, these standard constructions allow us to find a set of simple D-branes with specific properties. We give several concrete examples, which we use to illustrate the concepts of D-brane transport. 
\subsection{Constructions of matrix factorizations}
\label{sec-mfconstruction}
All the matrix factorizations that will be discussed here can be expressed in terms of a Clifford basis (\ref{clifford}). This will be enough for the purpose of determining a basis of (D0,D2,D4,D6)-branes for the monodromy calculations. For our GLSMs we are looking for matrix factorizations of the superpotential $W=pG_N(x_1,\ldots,x_5)$, where $G_N$ is a generic homogeneous polynomial. Since the hemisphere partition function is insensitive to deformations of the complex structure corresponding to deformations of the superpotential, we can restrict the parameters of $G_N$ to simplify the problem, provided that (\ref{regularity}) is still satisfied. Of particular interest is the Fermat point with
\begin{align}
  N&=6: & G_6^F&=x_{1}^{6}+x_{2}^{6}+x_{3}^{6}+x_{4}^{6}+x_{5}^{3}\\
  N&=8: & G_8^F&=x_{1}^{8}+x_{2}^{8}+x_{3}^{8}+x_{4}^{8}+x_{5}^{2}\\
  N&=10: & G_{10}^F&=x_{1}^{10}+x_{2}^{10}+x_{3}^{10}+x_{4}^{5}+x_{5}^{2}.
\end{align}
Matrix factorizations of Fermat-type superpotentials have been studied in the context of D-branes in B-type topological Landau-Ginzburg models, and we can make use of these results.

Let us start with matrix factorizations of the form
\begin{equation}
  \label{koszul}
  Q=\sum_i a_i\eta_i+b_i\bar{\eta}_i,\qquad \sum_{i}a_i\cdot b_i=W
  \end{equation}
One distinguishes whether the $a_i$ and $b_i$ are monomials or polynomials in the variables $x_1,\ldots,x_5$. Then one can find $32\times 32$ matrix factorizations $\bar{Q}$ of $G_N^F$ such that each monomial is factorized individually:
\begin{equation}
  \label{rsbrane}
  \bar{Q}=\sum_{i=1}^5 x_i^{k_i}\eta_i+x_i^{d_i-k_i}\bar{\eta}_i.
\end{equation}
In the context of Landau-Ginzburg orbifold theories, such matrix factorizations have been linked to Recknagel-Schomerus boundary states with label $L=(k_1-1,\ldots,k_5-1)$ \cite{Ashok:2004zb} of the associated Gepner model. They are also called tensor product branes, since they arise as tensor products of boundary states of the minimal model components of the CFT. The orbifold action defines $N$ matrices $\gamma_i$ via the condition $\gamma_i \bar{Q}(e^{2\pi i\frac{Q_i}{N}}x_i)\gamma_i^{-1}=Q(x_i)$. The data $\bar{Q}_i\equiv(\bar{Q},\gamma_i)$ with $i=1,\ldots,N$ defines $N$ branes in the $\mathbb{Z}_N$-orbit, where the extra label coincides with the label $M=(0,1,\ldots,N-1)$ of the boundary state.

To extend (\ref{rsbrane}) to a matrix factorization of the GLSM potential one has to add the $p$-field. There is a choice of whether one puts the $p$-field in front of the $\eta_i$-term or the $\bar{\eta}_i$ term for every monomial in $G_N^F$. Each choice leads to a different, not necessarily grade-restricted, brane in the GLSM. How to correctly lift each $\bar{Q}_i$ to grade-restricted GLSM brane is explained in sections 10.2.1 and 10.4.1 of \cite{Herbst:2008jq}. In the large volume phase matrix factorizations of this type map to objects which have a complicated structure in terms of (D0,D2,D4,D6)-charges. In the context of one-parameter hypersurfaces matrix factorizations of the type (\ref{rsbrane}) and their geometric counterparts have been discussed in \cite{Knapp:2008uw}.

If the Fermat polynomial of $G_N^F$ contains monomials of the form
\begin{equation}
  G^F_N(x_a,x_b)=x_a^{d_a}+x_b^{d_b}=x_a^{p r_a}+x_b^{p r_b},\qquad p=\mathrm{gcd}(d_a,d_b),
\end{equation}
one can find a factorization $G^F_N(x_a,x_b)=\prod_{m=0}^{p-1}(x_a^{r_1}-e^{\frac{2m+1}{p}\pi i}x_b^{r_2})$.
Now define a set $K=\{0,1,\ldots, p-1\}$ and a subset $I\subset K$. Then one can construct $2\times 2$ matrix factorizations 
\begin{equation}
  \label{permutation}
  \bar{Q}(x_a,x_b)=\prod_{m\in I}(x_a^{r_1}-e^{\frac{2m+1}{p}\pi i}x_b^{r_2})\eta+\prod_{m\in K\backslash I}(x_a^{r_1}-e^{\frac{2m+1}{p}\pi i}x_b^{r_2})\bar{\eta}.
  \end{equation}
A subset of matrix factorizations of this type has been identified with (generalized) permutation branes in the associated boundary CFT \cite{Brunner:2005fv,Enger:2005jk,Caviezel:2005th,Fredenhagen:2006qw}. To extend these two-variable matrix factorizations to matrix factorization of $G_N^F(x_1,\ldots,x_5)$ one can combine this with further permutation-type or tensor product-type factorizations to obtain a matrix factorization of $G_N^F$. Including the orbifold action again gives branes $\bar{Q}_i$, as above. By adding the $p$-field one obtains GLSM matrix factorizations of the form (\ref{koszul}), which are grade-restricted if one follows the algorithm described in \cite{Herbst:2008jq}. Since in our models $G_N^F$ has five monomials one can either get $8\times 8$ matrix factorizations where two pairs of monomials are factorized as in (\ref{permutation}), or  $16\times 16$ matrix factorizations where only one pair of monomials is factorized as in (\ref{permutation}) and the rest of the monomials is factorized individually. As we will show in the examples below and in section \ref{sec-monodromy}, one particular GLSM brane arising from the former construction corresponds to a D2-brane with minimal charge in the geometric phase. The second type of construction produces in particular a D0-brane with minimal charge. 

Aside from these CFT-inspired constructions of GLSM matrix factorizations, one can also construct branes from a geometric intuition. In the geometric phase we construct a B-type D-brane on the Calabi-Yau hypersurface by a complete intersection of the Calabi-Yau equation with a number of irreducible hypersurfaces given by equations $h_i$ in the ambient space
\begin{equation}
  G_N(x_1,\ldots,x_5)\cap\bigcap_{i=1}^{\delta}h_{i},
  \end{equation}
where for a number of $\delta=0,1,2,3$ hypersurfaces we get D6, D4, D2 and D0-branes respectively. These geometric branes can be lifted to the GLSM in the following way:
\begin{equation}
  \label{geom}
  Q=G_N\eta_0+p\bar{\eta}_0+\sum_{i=1}^{\delta}h_i\eta_i.
  \end{equation}
Note that the terms proportional to $\bar{\eta}_{i>0}$ have coefficient zero. The case $\delta=0$ describes a D6-brane that coincides with the empty brane (\ref{lgempty}) associated to the Landau-Ginzburg phase. GLSM branes such as (\ref{geom}) are never grade restricted. Viewing these matrix factorizations as GLSM lifts branes in the geometric phase, one first has to grade-restrict them using the empty brane associated to the large radius phase in order to make sense of them outside the geometric phase. 
\subsection{Examples}
Before we construct a basis of GLSM branes of which we compute the monodromy in section \ref{sec-monodromy}, let us give some examples of GLSM matrix factorizations and grade restriction. Every brane $\mathcal{B}$ describes a different object in the category of GLSM branes, unless there is an equivalence of matrix factorizations
\begin{equation}
  Q_2(\phi)=U(\phi)Q_{1}(\phi)U^{-1}(\phi),
  \end{equation}
where $U(\phi)$ is invertible with polynomial entries.

We are primarily dealing with the hemisphere partition function, which is insensitive to the explicit form of the entries in the matrix factorizations. This reflects the fact that the central charge only sees the K-theory, not the full category. In practice, this means that we can get the same brane factors for different objects $\mathcal{B}$, and as a result the same hemisphere partition function, irrespective of the choice of phase. For the quintic, for instance, we can arbitrarily permute the variables in the matrix factorizations and always get the same hemisphere partition function, even though the corresponding branes are different objects in the category. In the case where the chiral fields can have different $U(1)$-charges, one still has this, but one can also combine variables of different weights in matrix factorizations of the form (\ref{permutation}). Then it is slightly less obvious which branes have the same hemisphere partition function.

To see an example of this, let us focus on the case $N=10$ where there is the greatest variety of weights. With the vacuum charge chosen to be $q_{|0\rangle}=10$, $r_{|0\rangle}=5$ we can construct three matrix factorizations of the Fermat polynomial that are of the form
\begin{equation}
  \label{deg10ex1}
  Q_{D0,i}=f_i(x_k,x_l)\eta_0+pg_i(x_k,x_l)\bar{\eta}_0+\sum_{j=1}^3x_{m,j}\eta_j+px_{m,j}^{d_m-1}\bar{\eta}_j,
  \end{equation}
where by $x_{m,j}$ we label those three $x$-variables that to not appear in the polynomials $f_{\alpha}(x_k,x_l)$ and $g_{\alpha}(x_k,x_l)$. Now we consider the following three choices:
\begin{align}
\begin{aligned}
  \alpha=1:  &\quad& f_1&= x_{1} - e^{i \frac{\pi}{10}}  x_{2}&\qquad&  g_1=\prod\limits_{n=1}^{9} \left(x_{1} -e^{ \frac{2n+1}{10}\pi i }x_{2}\right), \\
  \alpha=2:  &\quad& f_2&=x_{1}^2 -e^{\frac{i \pi}{5}} x_{4} &\qquad & g_2=\prod\limits_{n=1}^{4} \left(x_{1}^{2} -e^{ \frac{2n+1}{5}\pi i }x_{4}\right), \\
  \alpha=3:  &\quad& f_3&= x_{1}^5 +i x_{5} &\qquad & g_3= x_{1}^5 -i x_{5}. \\
   \end{aligned}
  \end{align}
 As one easily can convince oneself, the Clifford matrices have the same gauge charges but get permuted:
\begin{align} 
\begin{aligned} 
\alpha&=1:\qquad &  q_{\eta_0}&=-1 & q_{\eta_1} &=-1 & q_{\eta_2} &=-2 & q_{\eta_3} &=-5 \\
\alpha&=2:\qquad &  q_{\eta_0}&= -2 & q_{\eta_1} &=-1 & q_{\eta_2} &=-1 & q_{\eta_3} &=-5 \\
\alpha&=3:\qquad &  q_{\eta_0}&= -5  & q_{\eta_1} &=-1 & q_{\eta_2} &=-1 & q_{\eta_3} &=-2 
\end{aligned} 
\end{align}
Therefore all three GLSM branes can be associated to a complex of Wilson line branes of the form
\begin{equation}
  \label{d10ex1complex}\mathcal{B}_{D0,\alpha}: \quad	
\xymatrix@1@C=5pt{\mathcal{W}(1)_{1}^{\oplus 1}\ar@<2pt>[rr]
	&&{\ar@<2pt>[ll]\begin{array}{c}
 \mathcal{W}(2)^{\oplus 2}_{2} \\
 \bigoplus \\
 \mathcal{W}(3)_{2}^{\oplus 1} \\
  \bigoplus \\
 \mathcal{W}(6)_{2}^{\oplus 1}
\end{array}}\ar@<2pt>[rr]
&&{\ar@<2pt>[ll]\begin{array}{c}
\mathcal{W}(3)_{3}^{\oplus 1} \\
 \bigoplus \\
 \mathcal{W}(4)^{\oplus 2}_{3} \\
  \bigoplus \\
 \mathcal{W}(7)^{\oplus 2}_{3} \\
  \bigoplus \\
 \mathcal{W}(8)_{3}^{\oplus 1}
\end{array}}\ar@<2pt>[rr]
&&{\ar@<2pt>[ll]\begin{array}{c}
 \mathcal{W}(5)_{4}^{\oplus 1} \\
  \bigoplus \\
 \mathcal{W}(8)_{4}^{\oplus 1} \\
  \bigoplus \\
 \mathcal{W}(9)^{\oplus 2}_{4}
\end{array}}\ar@<2pt>[rr]
&&\ar@<2pt>[ll] \mathcal{W}(10)_{5}^{\oplus 1}}.  
\end{equation}

The maps in this complex are different for the three branes. However, since their explicit form does not enter into the brane factor, we get
\begin{equation}
  f_{\mathcal{B}_{D0}}=-e^{2 \pi  \sigma }+2 e^{4 \pi  \sigma }-2 e^{8 \pi  \sigma }+e^{10 \pi  \sigma }+e^{12 \pi  \sigma }-2 e^{14 \pi  \sigma }+2 e^{18 \pi  \sigma }-e^{20 \pi  \sigma }
  \end{equation}
for all three cases. Evaluating the hemisphere partition function in the geometric phase we get
\begin{equation}
  Z^{\zeta \gg 0}_{D^{2}}(\mathcal{B}_{D0})= \varpi_{0},
\end{equation}
where the $\varpi_i$ are the periods (\ref{periods}) of the mirror Calabi-Yau and we have to replace $z=e^{-t}$. Hence, all three of these GLSM branes descend to D0-branes in $X_{10}:\mathbb{P}(11125)[10]$ that have the same charge but are described by different equations. Since (\ref{d10ex1complex}) is grade restricted with respect to the window $w_{1}:\, q\in\{1,2,\ldots,10\}$ associated to the path $\theta\in(-12\pi,-10\pi)$, evaluating the hemisphere partition function in the Landau-Ginzburg phase will give the charges of the branes of the analytic continuation of the three D0-branes into the Landau-Ginzburg phase along $w_1$.

An arbitrary GLSM brane is usually not grade restricted. As a further example consider the matrix factorization
\begin{equation}
  \label{deg10ex2}
  Q_{D2}=f_1(x_1,x_2)\eta_0+pg_1(x_1,x_2)\bar{\eta}_0+\tilde{f}_1(x_3,x_4)\eta_1+p\tilde{g}_2(x_3,x_4)\bar{\eta}_1+x_{5}\eta_2+px_{5}\bar{\eta}_2,
  \end{equation}
    with
    \begin{align}
\begin{aligned}
  f_{1} &=x_{1}^5 + i x_{2}^5 &\qquad & g_{1} =x_{1}^5 - i x_{2}^5,\\
  \tilde{f}_{1} &= x_{3}^2 -e^{\frac{i \pi}{5}} x_{4} &\qquad & \tilde{g}_{2} = \prod\limits_{n=1}^{4} \left(x_{3}^{2} -e^{ \frac{2n+1}{5}\pi i }x_{4}\right). 
\end{aligned}
    \end{align}
    
    With the same choice of vacuum charges as above, we can associate to this
\begin{equation}
  \label{d10ex2complex}
  	\mathcal{B}_{D2}: \quad \xymatrix@1@C=5pt{\mathcal{W}(-2)_{2}^{\oplus 1}\ar@<2pt>[rr]
	&&{\ar@<2pt>[ll] \begin{array}{c}
 \mathcal{W}(3)^{\oplus 2}_{3} \\
  \bigoplus \\
 \mathcal{W}(0)_{3}^{\oplus 1}
\end{array}}\ar@<2pt>[rr]
&&{\ar@<2pt>[ll] \begin{array}{c}
\mathcal{W}(8)_{4}^{\oplus 1} \\
 \bigoplus \\
 \mathcal{W}(5)^{\oplus 2}_{4} 
\end{array}}\ar@<2pt>[rr]
&&\ar@<2pt>[ll]  \mathcal{W}(10)_{5}^{\oplus 1}}.
  \end{equation}
This complex of Wilson line branes is not grade-restricted to any window. Nevertheless we can evaluate the hemisphere partition function in both phases. In the geometric phase we get
\begin{equation}
  \label{d10ex2z}
  Z^{\zeta\gg0}_{D^2}(\mathcal{B}_{D2})= -5\varpi_{0}+ 5 \varpi_{1}.
  \end{equation}
We have found an object which carries D0 and D2-charges in the geometric phase. The hemisphere partition function evaluated in the Landau-Ginzburg phase is also non-zero and therefore (\ref{d10ex2complex}) also descends to a non-trivial brane in the Landau-Ginzburg phase. The respective branes in the two phases are however {\em not} related via analytic continuation along a path in the K\"ahler moduli space. To make a sensible analytic continuation we have to grade-restrict. For example, if we want to analytically continue the geometric object with central charge (\ref{d10ex2z}) to the Landau-Ginzburg phase along the path $\theta\in(-12\pi,-10\pi)$, associated to the charge window $w_{1}$ we have to replace (\ref{d10ex2complex}) by a different GLSM brane which also satisfies (\ref{d10ex2z}). This is done by binding empty branes (\ref{lrempty}) associated to the large radius phase. Using the notation of \cite{Knapp:2016rec}, the procedure is summarized in the following table:
\begin{equation}
  \label{d10ex2lrgrr}
  \mathcal{B}_{D2}':\quad 
  \resizebox{0.8\textwidth}{!}{
 \begin{tabular}{cccccccc}
 &&&&&&&\#\\ 
 \toprule
 &\multirow{2}{*}{$\textcolor{teal}{\underline{\mathcal{W}(-2)^{\oplus1}_{2}}}$}&\textcolor{blue!80}{\udash{\udash{$\mathcal{W}(0)^{\oplus1}_{3}$}}}&$\mathcal{W}(5)^{\oplus2}_{4}$& \multirow{2}{*}{$\mathcal{W}(10)^{\oplus1}_{5}$}\\ 
&&$\mathcal{W}(3)^{\oplus2}_{3}$&$\mathcal{W}(8)^{\oplus1}_{4}$ & \\ 
 \midrule
 
 \multirow{4}{*}{$\textcolor{teal}{\underline{\mathcal{W}(-2)^{\oplus1}_{1}}}$}&$\textcolor{violet}{\underline{\underline{\mathcal{W}(-1)^{\oplus3}_{2}}}}$&$\textcolor{purple}{\udot{$\mathcal{W}(0)^{\oplus3}_{3}$}}$&$\mathcal{W}(1)^{\oplus1}_{4}$&$\mathcal{W}(3)^{\oplus1}_{5}$& \multirow{4}{*}{$\mathcal{W}(8)^{\oplus1}_{6}$}&&\multirow{4}{*}{1}\\ 
&\textcolor{blue!50}{\udot{\udot{$\mathcal{W}(0)^{\oplus1}_{2}$}}}&$\mathcal{W}(1)^{\oplus3}_{3}$&$\mathcal{W}(2)^{\oplus3}_{4}$&$\mathcal{W}(6)^{\oplus1}_{5}$ & \\ 
&$\mathcal{W}(3)^{\oplus1}_{2}$&$\mathcal{W}(4)^{\oplus3}_{3}$&$\mathcal{W}(5)^{\oplus3}_{4}$&$\mathcal{W}(7)^{\oplus3}_{5}$ & \\ 
& &$\mathcal{W}(5)^{\oplus1}_{3}$&$\mathcal{W}(6)^{\oplus3}_{4}$&  & \\
\\

&\multirow{4}{*}{\textcolor{blue!80}{\udash{\udash{$\mathcal{W}(0)^{\oplus1}_{2}$}}}}&$\mathcal{W}(1)^{\oplus3}_{3}$&$\mathcal{W}(2)^{\oplus3}_{4}$&$\mathcal{W}(3)^{\oplus1}_{5}$&$\mathcal{W}(5)^{\oplus1}_{6}$& \multirow{4}{*}{$\mathcal{W}(10)^{\oplus1}_{7}$} &\multirow{4}{*}{1}\\ 
&&$\mathcal{W}(2)^{\oplus1}_{3}$&$\mathcal{W}(3)^{\oplus3}_{4}$&$\mathcal{W}(4)^{\oplus3}_{5}$&$\mathcal{W}(8)^{\oplus1}_{6}$ & \\ 
&&$\mathcal{W}(5)^{\oplus1}_{3}$&$\mathcal{W}(6)^{\oplus3}_{4}$&$\mathcal{W}(7)^{\oplus3}_{5}$&$\mathcal{W}(9)^{\oplus3}_{6}$ & \\ 
&& &$\mathcal{W}(7)^{\oplus1}_{4}$&$\mathcal{W}(8)^{\oplus3}_{5}$&  & \\

\multirow{4}{*}{\textcolor{violet}{\underline{\underline{$\mathcal{W}(-1)^{\oplus1}_{1}$}}}}&\textcolor{red!50}{\udash{$\mathcal{W}(0)^{\oplus3}_{2}$}}&$\mathcal{W}(1)^{\oplus3}_{3}$&$\mathcal{W}(2)^{\oplus1}_{4}$&$\mathcal{W}(4)^{\oplus1}_{5}$& \multirow{4}{*}{$\mathcal{W}(9)^{\oplus1}_{6}$}&&\multirow{4}{*}{3}\\ 
&$\mathcal{W}(1)^{\oplus1}_{2}$&$\mathcal{W}(2)^{\oplus3}_{3}$&$\mathcal{W}(3)^{\oplus3}_{4}$&$\mathcal{W}(7)^{\oplus1}_{5}$ &  \\ 
&$\mathcal{W}(4)^{\oplus1}_{2}$&$\mathcal{W}(5)^{\oplus3}_{3}$&$\mathcal{W}(6)^{\oplus3}_{4}$&$\mathcal{W}(8)^{\oplus3}_{5}$ & \\ 
& &$\mathcal{W}(6)^{\oplus1}_{3}$&$\mathcal{W}(7)^{\oplus3}_{4}$&  & \\ 
\\
\multirow{4}{*}{\textcolor{blue!50}{\udot{\udot{$\mathcal{W}(0)^{\oplus1}_{1}$}}}}&$\mathcal{W}(1)^{\oplus3}_{2}$&$\mathcal{W}(2)^{\oplus3}_{3}$&$\mathcal{W}(3)^{\oplus1}_{4}$&$\mathcal{W}(5)^{\oplus1}_{5}$& \multirow{4}{*}{$\mathcal{W}(10)^{\oplus1}_{6}$} &&\multirow{4}{*}{1}\\ 
&$\mathcal{W}(2)^{\oplus1}_{2}$&$\mathcal{W}(3)^{\oplus3}_{3}$&$\mathcal{W}(4)^{\oplus3}_{4}$&$\mathcal{W}(8)^{\oplus1}_{5}$ & \\ 
&$\mathcal{W}(5)^{\oplus1}_{2}$&$\mathcal{W}(6)^{\oplus3}_{3}$&$\mathcal{W}(7)^{\oplus3}_{4}$&$\mathcal{W}(9)^{\oplus3}_{5}$ & \\ 
& &$\mathcal{W}(7)^{\oplus1}_{3}$&$\mathcal{W}(8)^{\oplus3}_{4}$&  & \\
\\
&\multirow{4}{*}{\textcolor{purple}{\udot{$\mathcal{W}(0)^{\oplus1}_{2}$}}}&$\mathcal{W}(1)^{\oplus3}_{3}$&$\mathcal{W}(2)^{\oplus3}_{4}$&$\mathcal{W}(3)^{\oplus1}_{5}$&$\mathcal{W}(5)^{\oplus1}_{6}$& \multirow{4}{*}{$\mathcal{W}(10)^{\oplus1}_{7}$} &\multirow{4}{*}{3}\\ 
&&$\mathcal{W}(2)^{\oplus1}_{3}$&$\mathcal{W}(3)^{\oplus3}_{4}$&$\mathcal{W}(4)^{\oplus3}_{5}$&$\mathcal{W}(8)^{\oplus1}_{6}$ & \\ 
&&$\mathcal{W}(5)^{\oplus1}_{3}$&$\mathcal{W}(6)^{\oplus3}_{4}$&$\mathcal{W}(7)^{\oplus3}_{5}$&$\mathcal{W}(9)^{\oplus3}_{6}$ & \\ 
&& &$\mathcal{W}(7)^{\oplus1}_{4}$&$\mathcal{W}(8)^{\oplus3}_{5}$&  & \\ 

\\
 \multirow{4}{*}{\textcolor{red!50}{\udash{$\mathcal{W}(0)^{\oplus1}_{1}$}}}&$\mathcal{W}(1)^{\oplus3}_{2}$&$\mathcal{W}(2)^{\oplus3}_{3}$&$\mathcal{W}(3)^{\oplus1}_{4}$&$\mathcal{W}(5)^{\oplus1}_{5}$& \multirow{4}{*}{$\mathcal{W}(10)^{\oplus1}_{6}$} &&\multirow{4}{*}{9}\\ 
&$\mathcal{W}(2)^{\oplus1}_{2}$&$\mathcal{W}(3)^{\oplus3}_{3}$&$\mathcal{W}(4)^{\oplus3}_{4}$&$\mathcal{W}(8)^{\oplus1}_{5}$ & \\ 
&$\mathcal{W}(5)^{\oplus1}_{2}$&$\mathcal{W}(6)^{\oplus3}_{3}$&$\mathcal{W}(7)^{\oplus3}_{4}$&$\mathcal{W}(9)^{\oplus3}_{5}$ & \\ 
& &$\mathcal{W}(7)^{\oplus1}_{3}$&$\mathcal{W}(8)^{\oplus3}_{4}$&  & \\ 
 \bottomrule
 \end{tabular}}
  \end{equation}
The colors/underlines indicate the positions of the trivial brane-antibrane pairs that can be removed, and the column $\#$ indicates the number of copies of the respective brane. The result is grade-restricted with respect to the desired charge window. 
The brane factor is
\begin{equation}
f_{\mathcal{B}'_{D2}}=5 e^{2 \pi  \sigma } \left(e^{2 \pi  \sigma }-1\right)^3 \left(-3 e^{2 \pi  \sigma }-2 e^{4 \pi  \sigma }-2 e^{6 \pi  \sigma }-2 e^{8 \pi  \sigma }+e^{12 \pi  \sigma }-2\right).
\end{equation}
One indeed finds $Z^{\zeta\gg0}_{D^2}(\mathcal{B}_{D2})=Z^{\zeta\gg0}_{D^2}(\mathcal{B}'_{D2})$, but the results in the Landau-Ginzburg phase are different. The low-energy branes coming from (\ref{d10ex2lrgrr}) are now related via analytic continuation. 
\section{Monodromy for $U(1)$ GLSMs}
\label{sec-monodromy}
In this section we explicitly compute the monodromy matrices for one-parameter Calabi-Yau hypersurfaces in weighted projected space via the hemisphere partition function (\ref{z1par}). 
At first we have to find a suitable basis of branes. We will choose this with respect to the large radius phase, where we take branes $(\mathcal{O}_{\mathrm{pt}},\mathcal{O}_{l},\mathcal{O}_{H},\mathcal{O}_X)$ that represent a point, a line, a hyperplane and the structure sheaf of the Calabi-Yau. These lift to GLSM branes  $(\mathcal{B}_{\mathrm{pt}},\mathcal{B}_{l},\mathcal{B}_{H},\mathcal{B}_X)$. Specifically, our basis is such that the hemisphere partition function in the large radius phase evaluates to
\begin{align}
\begin{aligned}
Z^{\zeta\gg 0}_{D^{2}} \left(\mathcal{B}_{pt}\right)&= \varpi_{0} \\
Z^{\zeta\gg 0}_{D^{2}} \left(\mathcal{B}_{l} \right)&= \varpi_{1} \\
Z^{\zeta\gg 0}_{D^{2}} \left(\mathcal{B}_{H} \right)&=  \left(\frac{c_{2}\cdot H}{24}+\frac{H^{3}}{6}\right)\varpi_{0} - \frac{H^{3}}{2} \varpi_{1} + \frac{H^{3}}{2} \varpi_{2} \\
Z^{\zeta\gg 0}_{D^{2}} \left(\mathcal{B}_{X} \right)&=  \frac{c_{3}\zeta(3)}{\left(2 \pi i\right)^{3}}
\varpi_{0} + \frac{c_{2} \cdot H}{24} \varpi_{1} + \frac{H^{3}}{6} \varpi_{3}, \\
\end{aligned}
  \label{basis}
\end{align}
where the topological data (\ref{topdata}) of the Calabi-Yau enters. 
This basis is natural from the point of view of the D-brane constructions discussed in section \ref{sec-onepar}. It does, however, not correspond to the standard symplectic basis of periods as introduced in \cite{Candelas:1993dm}. To show this, we can compute the annulus partition function \cite{Honda:2013uca,Hori:2013ika}
\begin{equation}
  \chi(\mathcal{B}_1,\mathcal{B}_2)=\frac{1}{|W_G|}\int_{\Gamma}\frac{d^{\mathrm{rk}G}}{(2\pi i)^{\mathrm{rk}G}}\frac{\prod_{\alpha>0}\left(2\sinh \left(\frac{\alpha(u)}{2}\right)\right)^2}{\prod_i2\sinh\left(\frac{Q_i(u)}{2}\right)}f_{\mathcal{B}_1}\left(-\frac{u}{2\pi}\right)f_{\mathcal{B}_2}\left(\frac{u}{2\pi}\right),
  \end{equation}
where $|W_G|$ is the rank of the Weyl group and we choose the contour $\Gamma$ such that it encircles $u=0$. This gives the open Witten index. For the GLSM branes satisfying (\ref{basis}) we find
\begin{equation}
  \chi(\mathcal{B}_i,\mathcal{B}_j):\qquad
\begin{array}{c|rrrr}
\mathcal{B}_i\backslash \mathcal{B}_j&\mathcal{B}_{\mathrm{pt}} &\mathcal{B}_{l} &\mathcal{B}_{H} &\mathcal{B}_X \\
\hline
\mathcal{B}_{\mathrm{pt}} &0&0&0&1\\
\mathcal{B}_{l} &0&0&-1&0\\
\mathcal{B}_{H} &0&1&0&\kappa\\
\mathcal{B}_X &-1&0&-\kappa&0
\end{array}
  \end{equation}
with $\kappa=4,4,3$ for $N=6,8,10$, respectively.

We choose our reference path in the moduli space by fixing the charge window
\begin{align}
 w_{0}&
 : &\theta &\in \left(-N \pi, \left(-N+2\right) \pi \right) & q &\in \left\{0,1,2,\dots,N-1 \right\}
\end{align}
All the GLSM matrix factorizations below have a representation in terms of a Clifford basis. In all examples we choose the gauge and R-charge of the vacuum $|0\rangle$ to be
\begin{align}
  \mathcal{B}_{pt} &,\mathcal{B}_{l}: & q_{|0\rangle}&=N-1 & r_{|0\rangle}&=5 \\
  \mathcal{B}_{H} &,\mathcal{B}_{X}: &   q_{|0\rangle}&=N & r_{|0\rangle}&=5. 
\end{align}
Using the recipes stated in section \ref{sec-branes} we can then compute monodromy matrices $M_{LR}$, $M_{LG}$ and $M_C$ around the large radius, Landau-Ginzburg and conifold point. In all cases we find the relations
\begin{equation}
  M_{LG}^N={\mathbf{1}}\qquad M_{C}\cdot {M_{LR}^{-1}}\cdot M_{LG}={\mathbf{1}}.
  \end{equation}
Comparing with the literature, we see that what we call $M_{LR}$ is actually the inverse of the large radius monodromy matrix. This is consistent with the fact that we make a window shift $w_0\rightarrow w_{1}$ for both, the large radius and the Landau-Ginzburg monodromy.

In \cite{Font:1992uk,Klemm:1992tx} mirror symmetry was used to calculate these monodromies. Comparing with \cite{Klemm:1992tx}, our results must match with matrices $S$, $A$ and $T$ for the monodromies around the large radius, Landau-Ginzburg, and conifold points, respectively. They satisfy $S^{-1}=(AT)^N$. Indeed, we can find an integer matrix $U$ such that 
\begin{align}
      \label{kt-relation}
    S &= U \left(M_{LR}^{-1}\right)^{N} U^{-1} &
    A &= UM_{LG} U^{-1} &
    T&= U M_{C} U^{-1}.
\end{align}
\subsection{$X_6$: $\mathbb{P}(11112)[6]$}
We begin by determining GLSM branes that satisfy (\ref{basis}). For the D0-brane we choose the following matrix factorization of the superpotential $W=p G_6^F(x_1,\ldots,x_5)$
\begin{equation}
  Q_{\mathcal{B}_{\mathrm{pt}}}=  f_{1} \eta_{1} + x_{3} \eta_{2} +x_{4} \eta_{3} + x_{5} \eta_{4} + p g_{1} \bar{\eta}_{1} +p x_{3}^{5} \bar{\eta}_{2} +
	p x_{4}^{5}\bar{\eta}_{3} + p x^{2}_{5} \bar{\eta}_{4},
\end{equation}
with
  \begin{align}
  \begin{aligned}
  f_{1} = x_{1} - e^{i \frac{\pi}{6}} x_{2} &\qquad&
	g_{1} =\prod\limits_{n=1}^{5} \left(x_{1} -e^{  \frac{2n+1}{6}\pi i}x_{2}\right).
	\end{aligned}
  \end{align}
This translates into the following complex of Wilson line branes
\begin{equation}
  \mathcal{B}_{\mathrm{pt}}:\quad    \xymatrix@1@C=5pt{
\mathcal{W}(0)^{\oplus  1}_{1}
 \ar@<2pt>[rr]
&&{\ar@<2pt>[ll]
\begin{array}{c}
\mathcal{W}(2)^{\oplus 1}_{2}\\
\bigoplus \\
\mathcal{W}(1)^{\oplus 3}_{2}\\
\end{array}
}\ar@<2pt>[rr]
 &&{\ar@<2pt>[ll]
\begin{array}{c}
\mathcal{W}( 3)^{\oplus 3}_{3}\\
\bigoplus \\
\mathcal{W}( 2)^{\oplus 3}_{3}\\
\end{array}
}\ar@<2pt>[rr]
&&{\ar@<2pt>[ll]
\begin{array}{c}
\mathcal{W}(4)^{\oplus 3}_{4}\\
\bigoplus \\
\mathcal{W}(3)^{\oplus 1}_{4}\\
\end{array}
}\ar@<2pt>[rr]
&&
\ar@<2pt>[ll] \mathcal{W}(5)^{\oplus 1}_{5}
}.
  \end{equation}
The brane is grade restricted to the desired charge-window and the brane factor is
\begin{equation}
  \label{6d0}
  f_{\mathcal{B}_{\mathrm{pt}}}=3 e^{2 \pi  \sigma }-2 e^{4 \pi  \sigma }-2 e^{6 \pi  \sigma }+3 e^{8 \pi  \sigma }-e^{10 \pi  \sigma }-1.
\end{equation}
The D2-brane is obtained by an analogous construction:
\begin{equation}
  Q_{\mathcal{B}_l}=f_{1} \eta_{1} + f_{2} \eta_{2} + x_{5} \eta_{4} + p g_{1} \bar{\eta}_{1} +p g_{2} \bar{\eta}_{2} +
	 p x^{2}_{5} \bar{\eta}_{4},
\end{equation}
where 
\begin{align}
\begin{aligned}
	f_{1} &= x_{1} -e^{i \frac{\pi}{6}} x_{2}&\qquad &	g_{1} =\prod\limits_{n=1}^{5} \left(x_{1} -e^{  \frac{2n+1}{6}\pi i}x_{2}\right),\\
	f_{2} &= x_{3} - e^{i \frac{\pi}{6}} x_{4}&\quad &	g_{2} = \prod\limits_{n=1}^{5} \left(x_{3} -e^{  \frac{2n+1}{6}\pi i}x_{4}\right)	.
\end{aligned}
	\end{align}
The associated complex of Wilson line branes is
\begin{equation}
  \mathcal{B}_{l}:\quad\xymatrix@1@C=5pt{
\mathcal{W}(1)^{\oplus  1}_{2}
 \ar@<2pt>[rr]
 &&\ar @<2pt>[ll]{
\begin{array}{c}
\mathcal{W}(3)^{\oplus 1}_{3}\\
\bigoplus \\
\mathcal{W}(2)^{\oplus 2}_{3}\\
\end{array}
}\ar@<2pt>[rr]
&&\ar @<2pt>[ll]{
\begin{array}{c}
\mathcal{W}(4)^{\oplus 2}_{4}\\
\bigoplus \\
\mathcal{W}(3)^{\oplus 1}_{4}\\
\end{array}
}\ar@<2pt>[rr]
&&\ar @<2pt>[ll]
\mathcal{W}(5)^{\oplus 1}_{5}
}
\end{equation}
This brane is also grade-restricted and the brane factor is
\begin{equation}
  \label{6d2}
  f_{\mathcal{B}_{l}}=e^{2 \pi  \sigma }-2 e^{4 \pi  \sigma }+2 e^{8 \pi  \sigma }-e^{10 \pi  \sigma }.
\end{equation}
To construct a D4-brane we intersect a hyperplane characterized by the equation $h=\sum_{i=1}^4\alpha_ix_i$ with the hypersurface equation $G_6(x_1,\ldots,x_5)$. The GLSM-lift of such a configuration is given by the matrix factorization
\begin{equation}
  \label{6d4nongr}
  \widehat{Q}_{H}=G_6 \eta_{0} + p \bar{\eta}_{0}+h \eta_{1}
\end{equation}
This brane is not grade-restricted with respect to any window, as one can read off from the associated complex of Wilson line branes:
\begin{equation}
  \widehat{\mathcal{B}}_H:\quad\xymatrix@1@C=5pt{\mathcal{W}(-1)_{3}\ar@<2pt>[rr] 
	&&\ar @<2pt>[ll]{\begin{array}{c}
 \mathcal{W}(0)_{4} \\
 \mathcal{W}(5)_{4} 
\end{array}}\ar@<2pt>[rr]
&&\ar @<2pt>[ll] \mathcal{W}(6)_{5}}.
  \end{equation}
The grade restriction procedure to the window $w_0$ is indicated in the following table.
\begin{equation}
  \mathcal{B}_H:\quad 
  \resizebox{0.8\textwidth}{!}{
 \begin{tabular}{cccccccccc}
 &&&&&&&&&\#\\ 
 \toprule
 &&&&\multirow{2}{*}{$\textcolor{violet}{\underline{\mathcal{W}(-1)^{\oplus 1}_{3}}}$} &$\mathcal{W}(5)^{\oplus 1}_{4}$ &\multirow{2}{*}{$\textcolor{teal}{\underline{\underline{\mathcal{W}(6)^{\oplus  1}_{5}}}}$} &&\\
 &&&&&$\mathcal{W}(0)^{\oplus 1}_{4}$& &&& \\
 \midrule
 &&&\multirow{2}{*}{$\textcolor{violet}{\underline{\mathcal{W}(-1)^{\oplus  1}_{2}}}$}&$\mathcal{W}(1)^{\oplus 1}_{3}$ &$\mathcal{W}(2)^{\oplus 4}_{4}$ &$\mathcal{W}(3)^{\oplus 6}_{5}$&$\mathcal{W}(4)^{\oplus 4}_{6}$ &\multirow{2}{*}{$\mathcal{W}(5)^{\oplus 1}_{7}$} &\multirow{2}{*}{1}\\
 &&&&$\mathcal{W}(0)^{\oplus 4}_{3}$ & $\mathcal{W}(1)^{\oplus 6}_{4}$ &$\mathcal{W}(2)^{\oplus 4}_{5}$ &$\mathcal{W}(3)^{\oplus 1}_{6}$  &&\\
 \\
 
 \multirow{2}{*}{$\mathcal{W}(0)^{\oplus  1}_{-1}$}&$\mathcal{W}(2)^{\oplus 1}_{0}$ &$\mathcal{W}(3)^{\oplus 4}_{1}$ &$\mathcal{W}(4)^{\oplus 6}_{2}$&$\mathcal{W}(5)^{\oplus 4}_{3}$ &\multirow{2}{*}{$\textcolor{teal}{\underline{\underline{\mathcal{W}(6)^{\oplus 1}_{4}}}}$} &&&&\multirow{2}{*}{1}\\
 &$\mathcal{W}(1)^{\oplus 4}_{0}$ & $\mathcal{W}(2)^{\oplus 6}_{1}$ &$\mathcal{W}(3)^{\oplus 4}_{2}$ &$\mathcal{W}(4)^{\oplus 1}_{3}$ &&&&&\\
 \bottomrule
 \end{tabular}}
  \end{equation}
The brane factor is
\begin{equation}
  \label{6d4}
  f_{\mathcal{B}_H}=9 e^{2 \pi  \sigma }-5 e^{4 \pi  \sigma }-5 e^{6 \pi  \sigma }+9 e^{8 \pi  \sigma }-4 e^{10 \pi  \sigma }-4.
\end{equation}
One easily confirms that evaluating (\ref{z1par}) with the brane factors associated to $\widehat{\mathcal{B}}_H$ and $\mathcal{B}_H$ in the large radius phase yields the same result. 

The GLSM matrix factorization associated to the structure sheaf in the geometric phase is
\begin{equation}
  \label{6d6nongr}
  \widehat{Q}_{X}= G \eta_{1} + p \bar{\eta}_{1}
\end{equation}
with
\begin{equation}
  \widehat{\mathcal{B}}_X:\quad \xymatrix@1@C=5pt{\mathcal{W}(0)_{4}\ar@<2pt>[rr]
	&&\ar @<2pt>[ll] \mathcal{W}(6)_{5}}.
\end{equation}
As observed, this is the empty brane (\ref{lgempty}) associated to the Landau-Ginzburg phase. To grade-restrict we bind one copy of the empty brane (\ref{lrempty}) of the large radius phase to obtain
\begin{equation}
  \label{6d6}
  \mathcal{B}_X:\quad \xymatrix@1@C=5pt{
\mathcal{W}(0)^{\oplus 1}_{-1} 
\ar@<2pt>[rr]
 &&\ar @<2pt>[ll]{\begin{array}{c}
 \mathcal{W}(2)^{\oplus   1}_{0} \\
 \bigoplus \\
 \mathcal{W}(1)^{\oplus    4}_{0}\\
\end{array}}\ar@<2pt>[rr]
&&\ar @<2pt>[ll] {\begin{array}{c}
 \mathcal{W}(3)^{\oplus   4}_{1} \\
 \bigoplus \\
  \mathcal{W}(2)^{\oplus    6}_{1}\\
\end{array}}\ar@<2pt>[rr]
&&\ar @<2pt>[ll] {\begin{array}{c}
 \mathcal{W}(4)^{\oplus  6}_{2} \\
 \bigoplus \\
  \mathcal{W}(3)^{\oplus  4}_{2} \\
\end{array}}\ar@<2pt>[rr]
&& \ar @<2pt>[ll]{\begin{array}{c}
 \mathcal{W}(5)^{\oplus 4}_{3} \\
 \bigoplus \\
  \mathcal{W}(4)^{\oplus 1}_{3} \\
\end{array}}\ar@<2pt>[rr]
&&\ar @<2pt>[ll]
\mathcal{W}(0)^{\oplus 1}_{4}
}. 
\end{equation}
The brane factor is
\begin{equation}
  f_{\mathcal{B}_X}=4 e^{2 \pi  \sigma }-5 e^{4 \pi  \sigma }+5 e^{8 \pi  \sigma }-4 e^{10 \pi  \sigma }.
  \end{equation}

Inserting the brane factors of (\ref{6d0}), (\ref{6d2}), (\ref{6d4}) and (\ref{6d6}) into (\ref{z1par}) and evaluating it in the large radius phase, we indeed find (\ref{basis}). 

Now we are ready to compute the monodromies. We start off with the large radius and Landau-Ginzburg monodromies for the D0-brane. Following the steps outlined in section \ref{sec-zd2monodromy} we obtain the following complex for the monodromy image of (\ref{6d0}):
\begin{equation}
  \mathcal{B}^{'LG/LR}_{\mathrm{pt}}:\quad \xymatrix@1@C=5pt{ 
\mathcal{W}(1)^{\oplus1}_{1}\ar@<2pt>[rr]
&&\ar @<2pt>[ll]{\begin{array}{c} 
\mathcal{W}(2)^{\oplus3}_{2}\\ 
\bigoplus \\ 
\mathcal{W}(3)^{\oplus1}_{2}\\ 
\end{array}} 
\ar@<2pt>[rr]
&&\ar @<2pt>[ll]{\begin{array}{c} 
\mathcal{W}(3)^{\oplus3}_{3}\\ 
\bigoplus \\ 
\mathcal{W}(4)^{\oplus3}_{3}\\ 
\end{array}} 
\ar@<2pt>[rr]
&&\ar @<2pt>[ll]{\begin{array}{c} 
\mathcal{W}(4)^{\oplus1}_{4}\\ 
\bigoplus \\ 
\mathcal{W}(5)^{\oplus3}_{4}\\ 
\end{array}} 
\ar@<2pt>[rr]
&& \ar @<2pt>[ll]\mathcal{W}(6)^{\oplus1}_{5}} 
\end{equation}
This is no longer grade-restricted to the desired charge window. In case of the large radius monodromy we use $\mathcal{B}_E^{\zeta\gg0}$ to go back to the original window:
\begin{equation}
  \widetilde{\mathcal{B}}^{',LR}_{\mathrm{pt}}:\quad 
  \resizebox{0.8\textwidth}{!}{
  \begin{tabular}{cccccccc}
  &&&&&&& \# \\
  \toprule
  &&\multirow{2}{*}{$\mathcal{W}(1)^{\oplus1}_{1}$}&$\mathcal{W}(2)^{\oplus3}_{2}$&$\mathcal{W}(3)^{\oplus3}_{3}$&$\mathcal{W}(4)^{\oplus1}_{4}$& \multirow{2}{*}{$\textcolor{teal}{\underline{\mathcal{W}(6)^{\oplus1}_{5}}}$}\\ 
&&&$\mathcal{W}(3)^{\oplus1}_{2}$&$\mathcal{W}(4)^{\oplus3}_{3}$&$\mathcal{W}(5)^{\oplus3}_{4}$ && \\ 
\midrule

\multirow{2}{*}{$\mathcal{W}(0)^{\oplus1}_{-1}$}&$\mathcal{W}(1)^{\oplus4}_{0}$&$\mathcal{W}(2)^{\oplus6}_{1}$&$\mathcal{W}(3)^{\oplus4}_{2}$&$\mathcal{W}(4)^{\oplus1}_{3}$& \multirow{2}{*}{$\textcolor{teal}{\underline{\mathcal{W}(6)^{\oplus1}_{4}}}$} && \multirow{2}{*}{1}\\ 
&$\mathcal{W}(2)^{\oplus1}_{0}$&$\mathcal{W}(3)^{\oplus4}_{1}$&$\mathcal{W}(4)^{\oplus6}_{2}$&$\mathcal{W}(5)^{\oplus4}_{3}$ & & \\ 
  \bottomrule
  \end{tabular}
  }
\end{equation}
This yields the same brane factor as the original brane (\ref{6d0}), reflecting the fact that the D0-brane does not transform under large radius monodromy. On the other hand, to describe the Landau-Ginzburg monodromy we use the empty brane (\ref{lgempty}) to grade-restrict to the charge window we started with. The result is
\begin{equation}
\widetilde{\mathcal{B}}^{',LG}_{\mathrm{pt}}:\quad 
\xymatrix@1@C=5pt{ 
\mathcal{W}(1)^{\oplus1}_{1}\ar@<2pt>[rr]
&&\ar @<2pt>[ll]{\begin{array}{c} 
\mathcal{W}(2)^{\oplus3}_{2}\\ 
\bigoplus \\ 
\mathcal{W}(3)^{\oplus1}_{2}\\ 
\end{array}} 
\ar@<2pt>[rr]
&&\ar @<2pt>[ll]{\begin{array}{c} 
\mathcal{W}(3)^{\oplus3}_{3}\\ 
\bigoplus \\ 
\mathcal{W}(4)^{\oplus3}_{3}\\ 
\end{array}} 
\ar@<2pt>[rr]
&&\ar @<2pt>[ll]{\begin{array}{c} 
\mathcal{W}(4)^{\oplus1}_{4}\\ 
\bigoplus \\ 
\mathcal{W}(5)^{\oplus3}_{4}\\ 
\end{array}} 
\ar@<2pt>[rr]
&& \ar @<2pt>[ll]\mathcal{W}(0)^{\oplus1}_{5}} 
\end{equation}
This yields the brane factor
\begin{align}
\begin{aligned}
  f_{\widetilde{\mathcal{B}}^{',LG}_{\mathrm{pt}}}&=-e^{2 \pi  \sigma }+3 e^{4 \pi  \sigma }-2 e^{6 \pi  \sigma }-2 e^{8 \pi  \sigma }+3 e^{10 \pi  \sigma }-1 \\
   &=f_{\mathcal{B}_{\mathrm{pt}}}-f_{\mathcal{B}_{X}}.
  \end{aligned} 
\end{align}
The numerical coefficients encode the monodromy behavior of the D0-brane around the Landau-Ginzburg point.

To compute the conifold monodromy we declare our reference point to be in the large radius phase. We will move the D0-brane along a clockwise path around the singular point sitting between $w_0$ and $w_1$. This corresponds to the situation depicted in figure \ref{fig-conifoldmonodromy}. We first use the empty brane (\ref{lgempty}) of the Landau-Ginzburg phase to grade-restrict to the window $w_{1}=\{1,2,\ldots,6\}$. This yields
\begin{equation}
  \mathcal{B}^{'C}_{\mathrm{pt}}:\quad 
\xymatrix@1@C=5pt{ 
\mathcal{W}(6)^{\oplus1}_{1}\ar@<2pt>[rr]
&&\ar @<2pt>[ll]{\begin{array}{c} 
\mathcal{W}(1)^{\oplus3}_{2}\\ 
\bigoplus \\ 
\mathcal{W}(2)^{\oplus1}_{2}\\ 
\end{array}} 
\ar@<2pt>[rr]
&&\ar @<2pt>[ll]{\begin{array}{c} 
\mathcal{W}(2)^{\oplus3}_{3}\\ 
\bigoplus \\ 
\mathcal{W}(3)^{\oplus3}_{3}\\ 
\end{array}} 
\ar@<2pt>[rr]
&&\ar @<2pt>[ll]{\begin{array}{c} 
\mathcal{W}(3)^{\oplus1}_{4}\\ 
\bigoplus \\ 
\mathcal{W}(4)^{\oplus3}_{4}\\ 
\end{array}} 
\ar@<2pt>[rr]
&&\ar @<2pt>[ll] \mathcal{W}(5)^{\oplus1}_{5}} 
\end{equation}
In the second step we use the empty brane of the large radius phase to grade restrict back to the window $w_{0}$. The procedure is summarized in the following table:
\begin{equation}
   \widetilde{\mathcal{B}}^{',C}_{\mathrm{pt}}:\quad
   \begin{tabular}{ccccccc}
   &&&&&&\#\\
   \toprule
   \multirow{2}{*}{$\textcolor{teal}{\underline{\mathcal{W}(6)^{\oplus1}_{1}}}$}&$\mathcal{W}(1)^{\oplus3}_{2}$&$\mathcal{W}(2)^{\oplus3}_{3}$&$\mathcal{W}(3)^{\oplus1}_{4}$& \multirow{2}{*}{$\mathcal{W}(5)^{\oplus1}_{5}$}\\ 
&$\mathcal{W}(2)^{\oplus1}_{2}$&$\mathcal{W}(3)^{\oplus3}_{3}$&$\mathcal{W}(4)^{\oplus3}_{4}$ & \\ 
\midrule
\\
\multirow{2}{*}{$\mathcal{W}(0)^{\oplus1}_{-5}$}&$\mathcal{W}(1)^{\oplus4}_{-4}$&$\mathcal{W}(2)^{\oplus6}_{-3}$&$\mathcal{W}(3)^{\oplus4}_{-2}$&$\mathcal{W}(4)^{\oplus1}_{-1}$& \multirow{2}{*}{$\textcolor{teal}{\underline{\mathcal{W}(6)^{\oplus1}_{0}}}$}\\ 
&$\mathcal{W}(2)^{\oplus1}_{-4}$&$\mathcal{W}(3)^{\oplus4}_{-3}$&$\mathcal{W}(4)^{\oplus6}_{-2}$&$\mathcal{W}(5)^{\oplus4}_{-1}$ & \\ 
 
\bottomrule
   \end{tabular}
  \end{equation}
From this we obtain the brane factor
\begin{align}
\begin{aligned}
  f_{\widetilde{\mathcal{B}}^{',C}_{\mathrm{pt}}}&=7 e^{2 \pi  \sigma }-7 e^{4 \pi  \sigma }-2 e^{6 \pi  \sigma }+8 e^{8 \pi  \sigma }-5 e^{10 \pi  \sigma }-1 \\
  &=f_{\mathcal{B}_{\mathrm{pt}}}+f_{\mathcal{B}_{X}}.
  \end{aligned}
\end{align}

The calculation for the other branes is completely analogous and we omit the details. From the relations $f_{\widetilde{\mathcal{B}}_i'}=\sum_jM_{ij}f_{\mathcal{B}_j}$  obtain the monodromy matrices $M_{LR}$, $M_{LG}$ and $M_{C}$. Our results are
\begin{align}
  M_{LR}&=\begin{pmatrix}
1 &0&0&0 \\
1&1&0&0 \\
0&3&1&0 \\
0&3&1&1  
\end{pmatrix}   
&
 M_{LG}&=
 \begin{pmatrix}
1&0&0&-1 \\
1&1&0&-1 \\
0&3&1&-4 \\
0&3&1&-3 
 \end{pmatrix}
&
 M_C&=
 \begin{pmatrix}
 1 & 0&0 &1 \\
 0&1&0&0 \\
 0&0&1&4 \\
 0&0&0&1
 \end{pmatrix}
\end{align}
Comparing with \cite{Klemm:1992tx}, our result matches via (\ref{kt-relation}) if we choose
\begin{equation}
  U=\begin{pmatrix}
  0&1&0&0 \\
  -1&0&0&0\\
  4&3&-1&0 \\
  0&0&0&-1
  \end{pmatrix}.
  \end{equation}
\subsection{$X_8$: $\mathbb{P}(11114)[8]$}
Since the calculation works the same way as for the degree six hypersurface, we only state the results. Examples of matrix factorizations for the D0- and D2-branes satisfying (\ref{basis}) are
\begin{align}
   Q_{\mathcal{B}_{\mathrm{pt}}}&= f_{1} \eta_{1} + x_{3} \eta_{2} +x_{4} \eta_{3} + x_{5} \eta_{4} + p g_{1} \bar{\eta}_{1} +p x_{3}^{7} \bar{\eta}_{2} +
	p x_{4}^{7}\bar{\eta}_{3} + p x_{5} \bar{\eta}_{4},\\
    Q_{\mathcal{B}_{\mathrm{l}}}&=\tilde{f}_{1} \eta_{1} + \tilde{f}_{2} \eta_{2} + x_{5} \eta_{4} + p \tilde{g}_{1} \bar{\eta}_{1} +p \tilde{g}_{2} \bar{\eta}_{2} +
	 p x_{5} \bar{\eta}_{4},
\end{align}
with
\begin{align}
\begin{aligned}
    f_{1} = x_{1} -e^{i \frac{\pi}{8}} x_{2} &\qquad & g_{1} = \prod\limits_{n=1}^{7} \left(x_{1}-e^{  \frac{2n+1}{8}\pi i}x_{2}\right)	, 
    \end{aligned}
\end{align}
and
    \begin{align}
    \begin{aligned}
    \tilde{f}_{1} &=x_{1} -e^{i \frac{\pi}{8}} x_{2} &\qquad & \tilde{g}_{1} =\prod\limits_{n=1}^{7} \left(x_{1} -e^{  \frac{2n+1}{8}\pi i}x_{2}\right),	\\
     \tilde{f}_{2} &=x_{3} -e^{i \frac{\pi}{8}} x_{4} &\qquad& \tilde{g}_{2}=\prod\limits_{n=1}^{7} \left(x_{3} -e^{  \frac{2n+1}{8}\pi i}x_{4}\right)	.
    \end{aligned}
    \end{align}
    They correspond to the following grade restricted branes
\begin{align}
\mathcal{B}_{\mathrm{pt}}&:\quad
\xymatrix@1@C=5pt{ 
\mathcal{W}(0)^{\oplus1}_{1}\ar@<2pt>[rr]
&&\ar @<2pt>[ll]{\begin{array}{c} 
\mathcal{W}(1)^{\oplus3}_{2}\\ 
\bigoplus \\ 
\mathcal{W}(4)^{\oplus1}_{2}\\ 
\end{array}} 
\ar@<2pt>[rr]
&&\ar @<2pt>[ll]{\begin{array}{c} 
\mathcal{W}(2)^{\oplus3}_{3}\\ 
\bigoplus \\ 
\mathcal{W}(5)^{\oplus3}_{3}\\ 
\end{array}} 
\ar@<2pt>[rr]
&&\ar @<2pt>[ll]{\begin{array}{c} 
\mathcal{W}(3)^{\oplus1}_{4}\\ 
\bigoplus \\ 
\mathcal{W}(6)^{\oplus3}_{4}\\ 
\end{array}} 
\ar@<2pt>[rr] 
&& \ar @<2pt>[ll]\mathcal{W}(7)^{\oplus1}_{5}} \\
 \mathcal{B}_{\mathrm{l}}&:\quad\xymatrix@1@C=5pt{ 
\mathcal{W}(1)^{\oplus1}_{2}\ar@<2pt>[rr]
&&\ar @<2pt>[ll]{\begin{array}{c} 
\mathcal{W}(2)^{\oplus2}_{3}\\ 
\bigoplus \\ 
\mathcal{W}(5)^{\oplus1}_{3}\\ 
\end{array}} 
\ar@<2pt>[rr]
&&\ar @<2pt>[ll]{\begin{array}{c} 
\mathcal{W}(3)^{\oplus1}_{4}\\ 
\bigoplus \\ 
\mathcal{W}(6)^{\oplus2}_{4}\\ 
\end{array}} 
\ar@<2pt>[rr]
&&\ar @<2pt>[ll] \mathcal{W}(7)^{\oplus1}_{5}} .
  \end{align}
The non-grade restricted GLSM lifts of the D4- and D6-branes are described by the same matrix factorizations (\ref{6d4nongr}) and (\ref{6d6nongr}) as for the degree six hypersurface with $G_6$ replaced by $G_8$. After grade-restriction we get the following branes
\begin{equation}
   \mathcal{B}_H:\quad
   \resizebox{0.8\textwidth}{!}{
   \begin{tabular}{cccccccccc}
 &&&&&&&&& \# \\ 
   \toprule
  &&&& \multirow{2}{*}{$\textcolor{violet}{\underline{\mathcal{W}(-1)^{\oplus1}_{3}}}$}&$\mathcal{W}(0)^{\oplus1}_{4}$& \multirow{2}{*}{$\textcolor{teal}{\underline{\underline{\mathcal{W}(8)^{\oplus1}_{5}}}}$} \\ 
&&&&&$\mathcal{W}(7)^{\oplus1}_{4}$ & \\ 
   \midrule
   
   \multirow{2}{*}{$\mathcal{W}(0)^{\oplus1}_{-1}$}&$\mathcal{W}(1)^{\oplus4}_{0}$&$\mathcal{W}(2)^{\oplus6}_{1}$&$\mathcal{W}(3)^{\oplus4}_{2}$&$\mathcal{W}(4)^{\oplus1}_{3}$& \multirow{2}{*}{$\textcolor{teal}{\underline{\underline{\mathcal{W}(8)^{\oplus1}_{4}}}}$}&&&&\multirow{2}{*}{1}\\ 
&$\mathcal{W}(4)^{\oplus1}_{0}$&$\mathcal{W}(5)^{\oplus4}_{1}$&$\mathcal{W}(6)^{\oplus6}_{2}$&$\mathcal{W}(7)^{\oplus4}_{3}$ & \\

&&&\multirow{2}{*}{$\textcolor{violet}{\underline{\mathcal{W}(-1)^{\oplus1}_{2}}}$}&$\mathcal{W}(0)^{\oplus4}_{3}$&$\mathcal{W}(1)^{\oplus6}_{4}$&$\mathcal{W}(2)^{\oplus4}_{5}$&$\mathcal{W}(3)^{\oplus1}_{6}$& \multirow{2}{*}{$\mathcal{W}(7)^{\oplus1}_{7}$}&\multirow{2}{*}{1}\\ 
&&&&$\mathcal{W}(3)^{\oplus1}_{3}$&$\mathcal{W}(4)^{\oplus4}_{4}$&$\mathcal{W}(5)^{\oplus6}_{5}$&$\mathcal{W}(6)^{\oplus4}_{6}$ & \\  
   \end{tabular}}
\end{equation}
and 
\begin{equation}
\mathcal{B}_{X}: \quad
  \xymatrix@1@C=5pt{ 
\mathcal{W}(0)^{\oplus1}_{-1}\ar@<2pt>[rr]
&&\ar @<2pt>[ll]{\begin{array}{c} 
\mathcal{W}(1)^{\oplus4}_{0}\\ 
\bigoplus \\ 
\mathcal{W}(4)^{\oplus1}_{0}\\ 
\end{array}} 
\ar@<2pt>[rr]
&&\ar @<2pt>[ll]{\begin{array}{c} 
\mathcal{W}(2)^{\oplus6}_{1}\\ 
\bigoplus \\ 
\mathcal{W}(5)^{\oplus4}_{1}\\ 
\end{array}} 
\ar@<2pt>[rr]
&&\ar @<2pt>[ll]{\begin{array}{c} 
\mathcal{W}(3)^{\oplus4}_{2}\\ 
\bigoplus \\ 
\mathcal{W}(6)^{\oplus6}_{2}\\ 
\end{array}} 
\ar@<2pt>[rr] 
&&\ar @<2pt>[ll]{\begin{array}{c} 
\mathcal{W}(4)^{\oplus1}_{3}\\ 
\bigoplus \\ 
\mathcal{W}(7)^{\oplus4}_{3}\\ 
\end{array}} 
\ar@<2pt>[rr]
&&\ar @<2pt>[ll] \mathcal{W}(0)^{\oplus1}_{4}} 
\end{equation}
Now we can perform the monodromy as discussed in section \ref{sec-zd2monodromy}. We find the following relations between the brane factors of the image branes $\widetilde{\mathcal{B}}'_i$ for the large radius, Landau-Ginzburg and conifold monodromies:
\begin{align}
  M_{LR}&=\begin{pmatrix}
1 &0&0&0 \\
1&1&0&0 \\
0&2&1&0 \\
0&2&1&1  
\end{pmatrix}   
&
 M_{LG}&=
 \begin{pmatrix}
1&0&0&-1 \\
1&1&0&-1 \\
0&2&1&-4 \\
0&2&1&-3 
 \end{pmatrix}
&
 M_C&=
 \begin{pmatrix}
 1 & 0&0 &1 \\
 0&1&0&0 \\
 0&0&1&4 \\
 0&0&0&1
 \end{pmatrix}
\end{align}
Comparing with \cite{Klemm:1992tx} we find a match if we choose the matrix $U$ in (\ref{kt-relation}) as follows
\begin{equation}
  U=
  \begin{pmatrix}
  0&1&0&0\\
  -1&0&0&0\\
  4&2&-1&0\\
  0&0&0&-1
  \end{pmatrix}.
  \end{equation}
\subsection{$X_{10}$: $\mathbb{P}(11125)[10]$}
Again the calculations are the same as for the previous hypersurfaces and therefore we will only summarize the results. 
Matrix factorizations representing a D0- and a D2-brane are given by
\begin{align}
   Q_{\mathcal{B}_{\mathrm{pt}}}&= f_{1} \eta_{1} + x_{3} \eta_{2} +x_{4} \eta_{3} + x_{5} \eta_{4} + p g_{1} \bar{\eta}_{1} +p x_{3}^{9} \bar{\eta}_{2} +
	p x_{4}^{4}\bar{\eta}_{3} + p x_{5} \bar{\eta}_{4},\\
    Q_{\mathcal{B}_{\mathrm{l}}}&=\tilde{f}_{1} \eta_{1} + \tilde{f}_{2} \eta_{2} + x_{5} \eta_{4} + p \tilde{g}_{1} \bar{\eta}_{1} +p \tilde{g}_{2} \bar{\eta}_{2} +
	 p x_{5} \bar{\eta}_{4},
\end{align}
with 
 \begin{align}
 \begin{aligned}
    f_{1} &= x_{1} -e^{i \frac{\pi}{10}} x_{2}&\qquad&
    g_{1} =\prod\limits_{n=1}^{9} \left(x_{1} -e^{  \frac{2n+1}{10}\pi i}x_{2}\right),
     \end{aligned}
 \end{align}
 and
     \begin{align}
     \begin{aligned}
     \tilde{f}_{1} &= x_{1} -e^{i \frac{\pi}{10}} x_{2}&\qquad &
    \tilde{g}_{1} = \prod\limits_{n=1}^{9} \left(x_{1} -e^{  \frac{2n+1}{10}\pi i}x_{2}\right), \\
    \tilde{f}_{2} &= x_{3}^2 -e^{\frac{\pi i}{5}} x_{4}&\qquad &
    \tilde{g}_{2} = \prod\limits_{n=1}^{4} \left(x_{3}^{2} -e^{  \frac{2n+1}{5}\pi i}x_{4}\right).
    \end{aligned}
    \end{align}

 These factorizations lead to the following grade restricted complexes
    \begin{align}
\mathcal{B}_{\mathrm{pt}}&:\quad\xymatrix@1@C=5pt{ 
\mathcal{W}(0)^{\oplus1}_{1}\ar@<2pt>[rr] 
&&\ar @<2pt>[ll]{\begin{array}{c} 
\mathcal{W}(1)^{\oplus2}_{2}\\ 
\bigoplus \\ 
\mathcal{W}(2)^{\oplus1}_{2}\\ 
\bigoplus \\ 
\mathcal{W}(5)^{\oplus1}_{2}\\
\end{array}} 
\ar@<2pt>[rr]
&&\ar @<2pt>[ll]{\begin{array}{c} 
\mathcal{W}(2)^{\oplus1}_{3}\\ 
\bigoplus \\ 
\mathcal{W}(3)^{\oplus2}_{3}\\ 
\bigoplus \\ 
\mathcal{W}(6)^{\oplus2}_{3}\\ 
\bigoplus \\ 
\mathcal{W}(7)^{\oplus1}_{3}\\ 
\end{array}} 
\ar@<2pt>[rr] 
&&\ar @<2pt>[ll]{\begin{array}{c} 
\mathcal{W}(4)^{\oplus1}_{4}\\ 
\bigoplus \\ 
\mathcal{W}(7)^{\oplus1}_{4}\\ 
\bigoplus \\ 
\mathcal{W}(8)^{\oplus2}_{4}\\ 
\end{array}} 
\ar@<2pt>[rr]
&&\ar @<2pt>[ll] \mathcal{W}(9)^{\oplus1}_{5}} 
\\
 \mathcal{B}_{\mathrm{l}}&: \quad\xymatrix@1@C=5pt{ 
\mathcal{W}(1)^{\oplus1}_{2}\ar@<2pt>[rr]
&&\ar @<2pt>[ll]{\begin{array}{c} 
\mathcal{W}(2)^{\oplus1}_{3}\\ 
\bigoplus \\ 
\mathcal{W}(3)^{\oplus1}_{3}\\ 
\bigoplus \\ 
\mathcal{W}(6)^{\oplus1}_{3}\\ 
\end{array}} 
\ar@<2pt>[rr]
&&\ar @<2pt>[ll]{\begin{array}{c} 
\mathcal{W}(4)^{\oplus1}_{4}\\ 
\bigoplus \\ 
\mathcal{W}(7)^{\oplus1}_{4}\\ 
\bigoplus \\ 
\mathcal{W}(8)^{\oplus1}_{4}\\ 
\end{array}} 
\ar@<2pt>[rr]
&& \ar @<2pt>[ll]\mathcal{W}(9)^{\oplus1}_{5}}  , 
  \end{align}
 D4- and D6-branes are constructed in the same way as in  (\ref{6d4nongr}) and (\ref{6d6nongr}), whereby  $G_6$  gets replaced by $G_{10}$ and $h=\sum_{i=1}^3\alpha_i x_i$. Grade-restriction leads to 
 the  following branes  
  \begin{equation}
   \mathcal{B}_H:\quad
   \resizebox{0.8\textwidth}{!}{
   \begin{tabular}{cccccccccc}
 &&&&&&&&& \# \\ 
   \toprule
  &&&&\multirow{2}{*}{$\textcolor{violet}{\underline{\mathcal{W}(-1)^{\oplus1}_{3}}}$}&$\mathcal{W}(0)^{\oplus1}_{4}$& \multirow{2}{*}{$\textcolor{teal}{\underline{\underline{\mathcal{W}(10)^{\oplus1}_{5}}}}$}\\ 
&&&&&$\mathcal{W}(9)^{\oplus1}_{4}$ & \\ 
   \midrule
   
   \multirow{4}{*}{$\mathcal{W}(0)^{\oplus1}_{-1}$}&$\mathcal{W}(1)^{\oplus3}_{0}$&$\mathcal{W}(2)^{\oplus3}_{1}$&$\mathcal{W}(3)^{\oplus1}_{2}$&$\mathcal{W}(5)^{\oplus1}_{3}$& \multirow{4}{*}{$\textcolor{teal}{\underline{\underline{\mathcal{W}(10)^{\oplus1}_{4}}}}$}&&&& \multirow{4}{*}{1}\\ 
&$\mathcal{W}(2)^{\oplus1}_{0}$&$\mathcal{W}(3)^{\oplus3}_{1}$&$\mathcal{W}(4)^{\oplus3}_{2}$&$\mathcal{W}(8)^{\oplus1}_{3}$ & \\ 
&$\mathcal{W}(5)^{\oplus1}_{0}$&$\mathcal{W}(6)^{\oplus3}_{1}$&$\mathcal{W}(7)^{\oplus3}_{2}$&$\mathcal{W}(9)^{\oplus3}_{3}$ & \\ 
& &$\mathcal{W}(7)^{\oplus1}_{1}$&$\mathcal{W}(8)^{\oplus3}_{2}$&  & \\ 

&&&\multirow{4}{*}{$\textcolor{violet}{\underline{\mathcal{W}(-1)^{\oplus1}_{2}}}$}&$\mathcal{W}(0)^{\oplus3}_{3}$&$\mathcal{W}(1)^{\oplus3}_{4}$&$\mathcal{W}(2)^{\oplus1}_{5}$&$\mathcal{W}(4)^{\oplus1}_{6}$& \multirow{4}{*}{$\mathcal{W}(9)^{\oplus1}_{7}$}&\multirow{4}{*}{1}\\ 
&&&&$\mathcal{W}(1)^{\oplus1}_{3}$&$\mathcal{W}(2)^{\oplus3}_{4}$&$\mathcal{W}(3)^{\oplus3}_{5}$&$\mathcal{W}(7)^{\oplus1}_{6}$ & \\ 
&&&&$\mathcal{W}(4)^{\oplus1}_{3}$&$\mathcal{W}(5)^{\oplus3}_{4}$&$\mathcal{W}(6)^{\oplus3}_{5}$&$\mathcal{W}(8)^{\oplus3}_{6}$ & \\ 
&&&& &$\mathcal{W}(6)^{\oplus1}_{4}$&$\mathcal{W}(7)^{\oplus3}_{5}$&  & \\ 
   \end{tabular}}
\end{equation}
and 
\begin{align} 
\mathcal{B}_{X}:\quad
 \xymatrix@1@C=5pt{ 
\mathcal{W}(0)^{\oplus1}_{-1}\ar@<2pt>[rr]
&&\ar @<2pt>[ll]{\begin{array}{c} 
\mathcal{W}(1)^{\oplus3}_{0}\\ 
\bigoplus \\ 
\mathcal{W}(2)^{\oplus1}_{0}\\ 
\bigoplus \\ 
\mathcal{W}(5)^{\oplus1}_{0}\\  
\end{array}} 
\ar@<2pt>[rr]
&&\ar @<2pt>[ll]{\begin{array}{c} 
\mathcal{W}(2)^{\oplus3}_{1}\\ 
\bigoplus \\ 
\mathcal{W}(3)^{\oplus3}_{1}\\ 
\bigoplus \\ 
\mathcal{W}(6)^{\oplus3}_{1}\\ 
\bigoplus \\ 
\mathcal{W}(7)^{\oplus1}_{1}\\ 
\end{array}} 
\ar@<2pt>[rr]
&&\ar @<2pt>[ll]{\begin{array}{c} 
\mathcal{W}(3)^{\oplus1}_{2}\\ 
\bigoplus \\ 
\mathcal{W}(4)^{\oplus3}_{2}\\ 
\bigoplus \\ 
\mathcal{W}(7)^{\oplus3}_{2}\\ 
\bigoplus \\ 
\mathcal{W}(8)^{\oplus3}_{2}\\ 
\end{array}} 
\ar@<2pt>[rr]
&&\ar @<2pt>[ll]{\begin{array}{c} 
\mathcal{W}(5)^{\oplus1}_{3}\\ 
\bigoplus \\ 
\mathcal{W}(8)^{\oplus1}_{3}\\ 
\bigoplus \\ 
\mathcal{W}(9)^{\oplus3}_{3}\\ 
\end{array}} 
\ar@<2pt>[rr]
&&\ar @<2pt>[ll] \mathcal{W}(0)^{\oplus1}_{4}} .
\end{align} 
The large radius, Landau-Ginzburg and conifold monodromies are 
calculated as described in  section \ref{sec-zd2monodromy}.
Performing the steps we find the following monodromy matrices:
\begin{align}
  M_{LR}&=\begin{pmatrix}
1 &0&0&0 \\
1&1&0&0 \\
0&1&1&0 \\
0&1&1&1  
\end{pmatrix}   
&
 M_{LG}&=
 \begin{pmatrix}
1&0&0&-1 \\
1&1&0&-1 \\
0&1&1&-3 \\
0&1&1&-2 
 \end{pmatrix}
&
 M_C&=
 \begin{pmatrix}
 1 & 0&0 &1 \\
 0&1&0&0 \\
 0&0&1&3 \\
 0&0&0&1
 \end{pmatrix}.
\end{align}
Matching our results with the results in \cite{Klemm:1992tx} using
equations  (\ref{kt-relation}) we find agreement for 
\begin{equation}
  U=
  \begin{pmatrix}
  3&0&-1&0\\
  -1&0&0&0\\
  0&-1&0&0\\
  0&0&0&-1
  \end{pmatrix}.
  \end{equation}
\section{Conclusions}
In this work we have shown how to compute monodromies for D-branes in one-parameter Calabi-Yaus using the GLSM and the hemisphere partition function. There are several possibilities for further research.

One straight-forward generalization is to consider examples of Calabi-Yaus with more than one K\"ahler parameter. In the case of abelian GLSMs the grade restriction-rule generalizes to a ``band restriction rule'' in a straight forward manner \cite{Herbst:2008jq}. Therefore D-brane transport is understood for this case and our approach to computing the monodromies should generalize smoothly. It would be worthwhile to study for instance two-parameter Calabi-Yaus hypersurfaces in weighted projected space and to recompute the monodromy matrices discussed in \cite{Candelas:1993dm,Candelas:1994hw} using the GLSM. One interesting aspect is that these GLSMs have hybrid phases. To compute the monodromy we need at least some understanding of branes in these phases. The minimum requirement is to identify the empty branes in those hybrid models, since we need them for grade restrictions. In this way we might learn some more about branes in hybrid models. 

A more profound generalization is to extend the discussion to non-abelian GLSMs. This in particular requires a generalization of \cite{Herbst:2008jq} to the non-abelian case. These models pose several challenges. The Calabi-Yaus arising from non-abelian GLSMs are no longer complete intersections in toric spaces. There are examples of complete intersections in more general ambient varieties such as Grassmannians.  Non-abelian GLSMs also lead to Calabi-Yaus that are determinantal varieties. These are connected to strongly coupled phases in the GLSM, which are hard to analyze. Due to these issues, D-branes in the phases of such models are much less studied than the D-branes in Landau-Ginzburg or geometric phases of abelian GLSMs. While they all must lift to matrix factorizations of the GLSM superpotential, it is not clear how to do this explicitly. Furthermore, the moduli spaces of Calabi-Yaus arising from non-abelian GLSMs have a more complicated structure. For instance, the one-parameter examples due to R{\o}dland \cite{rodland98} and Hosono-Takagi \cite{Hosono:2011np} whose GLSM description has been found in \cite{Hori:2006dk,Hori:2011pd} have three singular points at the phase boundary. This must be reflected in the grade-restriction rule in non-abelian GLSMs. Despite these complications, it seems that D-brane transport and in particular monodromy calculations as advertised here also work in the non-abelian case \cite{beijing,wip}. See also \cite{MR3223878,Addington:2014sla,Rennemo:2016oiu} for recent work in mathematics in this context.

Finally, the GLSM lends itself to a more categorical understanding of monodromies. While we made the point here, that we only need the brane factors of the hemisphere partition function to obtain the monodromy matrices, we actually have the full information about the concrete objects of the GLSM category. Thanks to \cite{Herbst:2008jq} we also know how they map to the corresponding objects in the D-brane categories associated to the Landau-Ginzburg and geometric phases. Therefore one can use the GLSM to study automorphisms in D-brane categories in explicit examples, as shown in \cite{MR2923950}. It would be interesting to extend this discussion, in particular to the case of non-abelian GLSMs.

\bibliographystyle{fullsort}
\bibliography{onepar-monodromy}
\end{document}